# "Technological Approach to Mind Everywhere (TAME): an experimentally-grounded framework for understanding diverse bodies and minds"

Michael Levin[1,2]

[1] Allen Discovery Center at Tufts University, Medford, MA, USA

[2] Wyss Institute for Biologically Inspired Engineering at Harvard University

**Running title:**  Life as it can be, Mind as it can be

**Abstract**

Synthetic biology and bioengineering provide the opportunity to create novel embodied cognitive systems (otherwise known as minds) in a very wide variety of chimeric architectures combining evolved and designed material and software. These advances are disrupting familiar concepts in the philosophy of mind, and require new ways of thinking about and comparing truly diverse intelligences, whose composition and origin are not like any of the available natural model species. In this Perspective, I introduce TAME - Technological Approach to Mind Everywhere - a framework for understanding and manipulating cognition in unconventional substrates. TAME formalizes a non-binary (continuous), empirically-based approach to strongly embodied agency. When applied to regenerating/developmental systems, TAME suggests a perspective on morphogenesis as an example of basal cognition. The deep symmetry between problem-solving in anatomical, physiological, transcriptional, and 3D (traditional behavioral) spaces drives specific hypotheses by which cognitive capacities can scale during evolution. An important medium exploited by evolution for joining active subunits into greater agents is developmental bioelectricity, implemented by pre-neural use of ion channels and gap junctions to scale cell-level feedback loops into anatomical homeostasis. This architecture of multi-scale competency of biological systems has important implications for plasticity of bodies and minds, greatly potentiating evolvability. Considering classical and recent data from the perspectives of computational science, evolutionary biology, and basal cognition, reveals a rich research program with many implications for cognitive science, evolutionary biology, regenerative medicine, and artificial intelligence.

# Introduction

All cognitive agents are collective intelligences, because we are all made of parts. There is no truly monadic, indivisible agent: all minds reside in physical systems made of components of various complexity. However, as human adults, our primary experience is that of a centralized, coherent Self which controls events in a top-down manner. That is also how we formulate models of learning ("the *rat* learned X"), moral responsibility, decision-making, and valence: at the center is a subject which has agency, serves as the locus of rewards and punishments, possesses (as a single functional unit) memories, exhibits preferences, and takes actions. And yet, under the hood, we find collections of cells which follow low-level rules via distributed, parallel functionality and give rise to emergent system-level dynamics. Much as single celled organisms transitioned to multicellularity during evolution, the single cells of an embryo construct *de novo,* and then operate, a unified Self during a single agent's lifetime. The compound agent supports memories, goals, and cognition that belongs to that Self and not to any of the parts alone. Thus, one of the most profound and far-reaching questions is that of scaling and unification: how do the activities of competent, lower-level agents give rise to a multiscale holobiont that is truly more than the sum of its parts? And, given the myriad of ways that parts can be assembled and relate to each other, is it possible to define ways in which truly diverse intelligences can be recognized, compared, and understood?

Here, I develop a framework to drive new theory and experiment in biology, cognition, evolution, and biotechnology from a multi-scale perspective on the nature and scaling of the cognitive Self. An important part of this research program is the need to encompass beings beyond the familiar conventional, evolved, static model animals with brains. The gaps in existing frameworks, and thus opportunities for fundamental advances, are revealed by a focus on plasticity of existing forms, and the functional diversity enabled by chimeric bioengineering. To illustrate how this framework can be applied to unconventional substrates, I explore a deep symmetry between behavior and morphogenesis, deriving hypotheses for dynamics that up- and down-scale Selves within developmental and phylogenetic timeframes, and at the same time strongly impact the speed and course of the evolutionary process. Anatomical homeostasis, as the result of the behavior of the swarm intelligence of cells provides a rich example of how an inclusive, forward-looking technological framework can connect philosophical questions with specific empirical research programs.

The philosophical context for the following perspective is summarized in (Table 1), and links tightly to the field of basal cognition via its fundamentally gradualist approach. It should be noted that the specific proposals for biological mechanisms that scale functional capacity are synergistic with, but not linearly dependent on, this conceptual basis. The hypotheses about how bioelectric networks scale cell computation into anatomical homeostasis, and the evolutionary dynamics of multi-scale competency, can be explored without accepting the "minds everywhere" commitments of the framework. However, together they form a coherent lens onto the life sciences which helps generate testable new hypotheses and integrate data from several subfields.

For the purposes of this paper, "cognition" refers not only to complex, self-reflexive advanced cognition or metacognition, but also recognizes many diverse capacities for adaptive responsiveness and actions of different levels of sophistication in conventional (evolved) life forms as well as bioengineered ones [1; 2; 3; 4] (Figure 1). The framework, TAME – Technological Approach to Mind Everywhere – adopts a practical, constructive engineering perspective on the optimal place for a given system on the continuum of cognitive sophistication. This gives rise to an axis of *persuadability* (Figure 2), which is closely related to the Intentional Stance [5] but made more explicit in terms of functional engineering approaches needed to implement prediction and control in practice. Persuadabilty refers to the type of conceptual and practical tools that are optimal to rationally modify a given system's behavior. The origin story (designed vs. evolved), composition, and other aspects are not definitive guides to the correct level of agency for a living or non-living system. Instead, one must perform experiments to see which kind of intervention strategy provides the most efficient prediction and control (thus, one aim should be generalizing the human-focused Turing Test and other IQ metrics into a broader agency detection toolkit, which perhaps could itself be implemented by a useful algorithm).

Our capacity to find new ways to understand and manipulate complex systems is strongly related to how we categorize agency in our world. Newton didn't invent two terms – gravity (for terrestrial objects falling) and perhaps *shmavity* (for the moon) – because it would have lost out on the much more powerful unification. TAME proposes a conceptual unification that would facilitate porting of tools across disciplines and model systems. We should avoid quotes around mental terms because there is no absolute, binary distinction between *it knows* and *it "knows"* – only a difference in the degree to which a model will be useful that incorporates such components.

Given this perspective, below I develop hypotheses about invariants that unify otherwise disparate-seeming problems, such as morphogenesis, behavior, and physiological allostasis. These hypotheses suggest a specific way to understand the scaling of cognitive capacity through evolution, make interesting predictions, and suggest novel experimental work. They also provide ways to think about the impending expansion of the "space of possible bodies and minds" via the efforts of bioengineers, which is sure to disrupt categories and conclusions that have been formed in the context of today's natural biosphere.

What of consciousness? It is likely impossible to understand consciousness without understanding cognition, and the emphasis of this paper is on testable, empirical impacts of ways to understand cognition in all of its guises. Thus, the focus of most of the discussion below is on cognitive function, not on phenomenal consciousness (in the sense of the "Hard Problem" [6]). The goal is to help advance and delineate an exciting emerging field at the intersection of biology, philosophy, and the information sciences. By proposing a new framework and examining it in a broad context of now physically realizable (not merely logically possible) living structures, it may be possible to bring conceptual, philosophical thought up to date with recent advances in science and technology. At stake are current knowledge gaps in evolutionary, developmental, and cell biology, a new roadmap for regenerative medicine, lessons that could be ported to artificial intelligence and robotics, and broader implications for ethics. I return to the subject of consciousness at the end, to sketch some implications of the TAME framework for efforts to address consciousness *per se*.

**Cognition: changing the Subject**

Even advanced animals are really collective intelligences [7; 8; 9], exploiting still poorly-understood scaling and binding features of metazoan architectures that share a continuum with looser swarms that have been termed "liquid brains" [10]. Studies of 'centralized control' focus on a brain, which is in effect a network of cells performing functions that many cell types, including bacteria, can do [11]. The embodied nature of cognition means that mental Selves are dependent on a highly plastic material substrate which changes not only on evolutionary time scales but also during the lifetime of the agent itself.

The central consequence of the composite nature of all intelligences is that the Self is subject to significant change in real-time (Figure 3). This means both slow maturation through experience (a kind of "software" change that doesn't disrupt traditional ways of thinking about agency), as well as radical changes of the material in which a given mind is implemented [12]. The owner, or subject of memories, preferences, and in more advanced cases, credit and blame, is very malleable. At the same time, fascinating mechanisms somehow ensure the persistence of Self (such as complex memories) despite drastic alterations of substrate. For example, the massive self-destruction of the caterpillar brain, followed by the morphogenesis of an entirely different brain suitable for the moth or beetle, does not wipe all the memories of the larva but somehow maps them onto behavioral capacities in the post-metamorphosis host, despite its entirely different body [13; 14; 15; 16; 17; 18]. Not only that, but memories can apparently persist following the complete regeneration of brains in some organisms [19; 20; 21] such as planaria, in which prior knowledge and behavioral tendencies are somehow transferred onto a newly-constructed brain.

This poorly-understood phenomenon highlights the fascinating plasticity of the body, brain, and mind, and makes it clear that traditional model systems in which cognition is mapped onto a stable, discrete, mature brain are insufficient to fully understand the relationship between the Self and its material substrate. Many scientists study the behavioral properties of caterpillars, and of butterflies, but the transition zone in-between, from the perspective of philosophy of mind and cognitive science, provides an important opportunity to study the mind-body relationship by changing the body during the lifetime of the agent (not just during evolution). Note that continuity of being across drastic biological remodeling is not only relevant for unusual cases in the animal kingdom, but is a fundamental property of most life – even humans change from a collection of cells to a functional individual, via a gradual morphogenetic process that constructs an active Self in real time. This has not been addressed in biology, and likewise not yet in computer science, where machine learning approaches use static neural networks (there is not a formalism for altering artificial neural networks' architecture on the fly).

What are the invariants that enable a Self to persist (and be recognizable by third-person investigations) despite such change? Memory is a good candidate [22; 23] (Figure 3). However, memories can be transferred between individuals, by transplants of brain tissue or molecular engrams [24; 25; 26; 27; 28; 29]. Importantly, the movement of memories across individual animals is only a special case of the movement of memory in biological tissue in general. Even when housed in the same "body", memories must move between tissues – for example, in a trained planarian's tail fragment re-imprinting its learned information onto the newly regenerated brain, or the movement of memories onto new brain tissue during metamorphosis. In addition to the spatial movement and re-mapping of memories onto new substrates, there is also a temporal component, as each memory is really an instance of communication between past and future Selves. The plasticity of biological bodies, made of cells that die, are born, and significantly rearrange their tissue architecture, suggest that the understanding of cognition is fundamentally a problem of collective intelligence: to understand how stable cognitive structures can persist and are mapped onto swarm *dynamics*, not stable structures.

This is applicable even to such "stable" forms as a human brain, which are often spoken of as a single Subject of experience and thought. First, the gulf between planarian regeneration / insect metamorphosis, and human brains, is going to be bridged by emerging therapeutics. It is inevitable that stem cell therapies for degenerative brain diseases [30; 31; 32] will confront us with humans whose brains are partially replaced by the naïve progeny of cells that were not present during the formation of memories and personality traits in the patient. Even prior to these advances, it was clear that phenomena such as multiple personality disorder [33], communication with non-verbal brain hemispheres in commissurotomy patients [34; 35], conjoined twins with fused brains [36; 37], etc. place human cognition onto a continuous spectrum with respect to the plasticity of integrated Selves that reside within a particular biological tissue implementation.

Importantly, animal model systems are now providing the ability to harness that plasticity for functional investigations of the body-mind relationship. For example, it is now easy to radically modify bodies in a time-scale that is much faster than evolutionary change, to study the inherent plasticity of minds without eons of selection to shape them to fit specific body architectures. When tadpoles are created to have eyes on their tails, instead of their heads, they are still readily able to perform visual learning tasks [38; 39]. Planaria can readily be made with two (or more) brains in the same body [40; 41], and human patients are now routinely augmented with novel inputs (such as sensory substitution [42; 43; 44; 45]) or novel effectors, such as instrumentized interfaces allowing thought to control engineered devices such as wheelchairs in addition to the default muscle-driven peripherals of their own bodies [46; 47; 48]. The central phenomenon here is plasticity: minds are not tightly bound to one specific underlying architecture (as most of our software is today), but readily mold to changes of genomic defaults. The logical extension of this progress is a focus on self-modifying living beings and the creation of new agents in which the mind:body system is simplified by entirely replacing one side of the equation with an engineered construct. The benefit would be that at least one half of the system is now well-understood.

For example, in hybrots, animal brains are functionally connected to robotics instead of their normal body [49; 50; 51; 52]. It doesn't even have to be an entire brain - a plate of neurons can learn to fly a flight simulator, and it lives in a new virtual world [53; 54; 55], as seen from the development of closed-loop neurobiological platforms [56; 57; 58; 59; 60; 61]. These kinds of results are reminiscent of Philosophy 101's "brain in a vat" experiment [62]. Brains adjust to driving robots and other devices as easily as they adjust to controlling a typical, or highly altered, living body because minds are somehow adapted and prepared to deal with body alterations – throughout development, metamorphosis and regeneration, and evolutionary change.

The massive plasticity of bodies, brains, and minds means that a mature cognitive science cannot just concern itself with understanding standard "model animals" as they exist right now. The typical "subject", such as a rat or fruit fly, which remains constant during the course of one's studies and is conveniently abstracted as a singular Self or intelligence, obscures the bigger picture. The future of this field must expand to frameworks that can handle all of the possible minds across an immense option space of bodies. Advances in bioengineering and artificial intelligence suggest that we or our descendants will be living in a world in which Darwin's "endless forms most beautiful" (this Earth's N=1 ecosystem outputs) are just a tiny sample of the true variety of possible beings. Biobots, hybrots, cyborgs, synthetic and chimeric animals, genetically and cellularly bioengineered living forms, humans instrumentized to knowledge platforms, devices, and each other – these technologies are going to generate beings whose body architectures are nothing like our familiar phylogeny. They will be a functional mix of evolved and designed components; at all levels, smart materials, software-level systems, and living tissue will be integrated into novel beings which function in their own exotic Umwelt. Importantly, the information that is used to specify such beings' form and function is no longer genetic – it is truly "epigenetic" because it comes not only from the creature's own genome but also from human minds (and eventually, robotic machine-learning-driven platforms) that use cell-level bioengineering to generate novel bodies from genetically wild-type cells. In these cases, the genetics are no guide to the outcome (which highlights some of the profound reasons that genetics is hard to use to truly predict cognitive form and function even in traditional living species).

Now is the time to begin to develop ways of thinking about truly novel bodies and minds, because the technology is advancing more rapidly than philosophical progress. Many of the standard philosophical puzzles – concerning brain hemisphere transplants, moving memories, replacing body/brain parts, etc. are now eminently doable in practice, while the theory of how to interpret the results lags. We now have the opportunity to begin to develop conceptual approaches to 1) understand beings without a convenient evolutionary back-story as explanations for their cognitive capacities (whose minds are created de-novo, and not shaped by long selection pressures toward specific capabilities), and 2) develop ways to analyze novel Selves that are not amenable to simple comparisons with related beings, not informed by their phylogenetic position relative to known standard species, and not predictable from an analysis of their genetics. The implications range across insights into evolutionary developmental biology, advancing bioengineering and artificial life research, new roadmaps for regenerative medicine, ability to recognize exobiological life, and the development of ethics for relating to novel beings whose composition offers no familiar phylogenetic touchstone. Thus, here I propose the beginnings of a framework designed to drive empirical research and conceptual/philosophical analysis that will be broadly applicable to minds regardless of their origin story or internal architecture.

**TAME: a proposal for a framework**

The Technological Approach to Mind Everywhere (TAME) framework seeks to establish a way to recognize, study, and compare truly diverse intelligences in the space of possible agents. The goal of this project is to identify deep invariants between cognitive systems of very different types of agents, and abstract away from inessential features such as composition or origin, which were sufficient heuristics with which to recognize agency in prior decades but will surely be insufficient in the future [63]. To flesh out this approach, I first make explicit some of its philosophical foundations, and then discuss specific conceptual tools that have been developed to begin the task of understanding embodied cognition in the space of mind-as-it-can-be (a sister concept to Langton's motto for the artificial life community – "life as it can be") [64].

*Philosophical foundations of an approach to diverse intelligences*

One key commitment of this research program is that of gradualism with respect to almost all important cognition-related properties: advanced minds are in important ways generated in a continuous manner from much more humble proto-cognitive systems. On this view, it is hopeless to look for a clear bright line that demarcates "true" cognition (such as that of humans, great apes, etc.) from metaphorical "as if cognition" or "just physics". Taking evolutionary biology seriously means that there is a continuous series of forms that connect any cognitive system with much more humble ones. While phylogenetic history already refutes views of a magical arrival of "true cognition" in across one generation, from parents that didn't have it (instead stretching the process of cognitive expansion over long time scales and slow modification), recent advances in biotechnology make this completely implausible. <u>For any putative difference between a creature that is proposed to have *true* preferences, memories, and plans and one that supposedly has *none*, we can now construct in-between, hybrid forms which then make it impossible to say whether the resulting being is an Agent or not</u>. Many pseudo-problems evaporate when a binary view of cognition is dissolved by an appreciation of the plasticity, and interoperability of living material at all scales of organization. A definitive discussion of the engineering of preferences and goal-directedness, in terms of hierarchy requirements and upper-directedness, is given in [65; 66].

For example, one view is that only biological, evolved forms have intrinsic motivation, while software AI agents are only faking it via functional performance (but don't actually *care* [67; 68; 69]). But which biological systems really care – fish? Single cells? Do mitochondria (which used to be independent organisms) have true preferences about physiological states? The lack of consensus on this question in purely biological systems highlights the futility of binary categories. Moreover, we can now readily construct hybrid systems which have any percentage of robotics tightly coupled to on-board living cells and tissues, which function together as one integrated being. How many living cells does a robot need to contain before the living system's "true" cognition bleeds over into the whole? On the continuum between human brains (with electrodes and a machine learning converter chip) that drive assistive devices (e.g., 95% human, 5% robotics), and robots with on-board cultured human brain cells instrumentized to assist with performance (5% human, 95% robotics), where can one draw the line – given that any desired percent combination is possible to make? No quantitative answer is sufficient to push a system "over the line" because there is no such line. Interesting aspect of agency or cognition are rarely Boolean values. Instead of a binary dichotomy, which leads to impassable philosophical roadblocks, we envision a continuum of phylogenetic advancement in information-processing capacity which has phase transitions that ramp up novel capabilities but is nevertheless a continuous process that is not devoid of proto-cognitive capacity before complex brains appear. This kind of gradualist view has been expounded in the context of evolutionary forces controlling individuality [70; 71], but here the focus is on events taking place within the lifetime of individuals and driven by information and control dynamics. The TAME framework asks "how much" and "what kind of" cognition any given system might manifest if we interacted with it in the right way, at the right scale of observation. And of course, the degree of cognition is not a single parameter that gives rise to a *scala naturae* but a shorthand for the shape and size of its cognitive capacities in a rich space (discussed below).

The second pillar of TAME is that there is no privileged material substrate for Selves. Alongside familiar materials such as brains made of neurons, the field of basal cognition [1; 2; 72; 73; 74; 75; 76; 77; 78; 79] has been identifying novel kinds of intelligences in single cells, plants, animal tissues, and swarms. Together with the fields of active matter, intelligent materials, swarm robotics, machine learning, and someday, exobiology, it is clear that we cannot rely on a familiar signature of "big vertebrate brain" as a necessary condition for mind. Molecular phylogeny shows that the specific components of brains pre-date the evolution of neurons per se, and life has been solving problems long before brains came onto the scene [80; 81; 82; 83; 84]. We must develop tools to characterize and relate to a wide diversity of minds in unconventional material implementations [85; 86; 87; 88; 89; 90; 91]. Closely related to that is the de-throning of natural evolution as the only acceptable origin story for a true Agent (many have proposed a distinction between evolved living forms vs. the somehow inadequate machines which were merely designed by man [63]). First, synthetic evolutionary processes are now being used in the lab to create "machines" and modify life [92; 93]. Second, the whole process of evolution, basically a hill-climbing search algorithm, is a set of frozen accidents and meandering selection among random tweaks. If this short-sighted process can give rise to true minds, then so can a rational engineering approach. There is nothing magical about evolution (driven by randomizing processes) as a forge for cognition; surely we can eventually do at least as well, and likely much better, using rational construction principles and an even wider range of materials.

The third foundational aspect of TAME is that the correct answer to how much agency a system has cannot be settled by philosophy – it is an empirical question. The goal is to produce a framework that drives experimental research programs, not only philosophical debate about what should or should not be possible as a matter of definition. To this end, the productive way to think about this a variant of Dennett's "intentional stance" [94; 95], which frames properties such as cognition as observer-dependent, empirically testable, and defined by how much benefit their recognition offers to science (Figure 2). Thus, the correct level of agency with which to treat any system must be determined by experiments that reveal which kind of model and strategy provides the most efficient predictive and control capability over the system. In this engineering (understand, modify, build)-centered view, the optimal position of a system on the spectrum of agency is determined empirically, based on which kind of model affords the most efficient way of prediction and control.

A standard methodology in science is to avoid attributing agency to a given system unless absolutely necessary. The mainstream view (e.g., Morgan's Canon) is that it's too easy to fall into a trap of "anthropomorphizing" systems with only apparent cognitive powers, when one should only be looking for models focused on mechanistic, lower levels of description that eschew any kind of teleology or mental capacity [96; 97]. However, analysis shows that this view provides no useful parsimony [98]. The rich history of debates on reductionism and mechanism needs to be complemented with an empirical, engineering approach that is not inappropriately slanted in one direction on this continuum. Teleophobia leads to Type 2 errors with respect to attribution of cognition that carry a huge opportunity cost for not only practical outcomes like regenerative medicine [99] and engineering, but also ethics. Humans (and many other animals) readily attribute agency to systems in their environment; scientists should be comfortable with testing out a theory of mind regarding various complex systems for the exact same reason - it can often greatly enhance prediction and control, by recognizing the true features of the systems with which we interact. This perspective implies that there is no such thing as "anthropomorphizing" because human beings have no unique essential property which can be inappropriately attributed to agents that have *none* of it. Instead, we should seek ways to naturalize human capacities as elaborations of more fundamental principles that are widely present in complex systems, in very different types and degrees, and to identify the *correct* level for any given system.

Avoiding philosophical wrangling over privileged levels of explanation [100; 101; 102], TAME takes an empirical approach to attributing agency, which increases the toolkit of ways to relate to complex systems, and also works to reduce profligate attributions of mental qualities. We do not say that a thermos knows whether to keep something hot or cold, because no model of thermos cognition does better than basic thermodynamics to explain its behavior or build better thermoses. At the same time, we know we cannot simply use Newton's laws to predict the motion of a (living) mouse at the top of a hill, requiring us to construct models of navigation and goal-directed activity for the controller of the mouse's behavior over time [103].

Under-estimating the capacity of a system for plasticity, learning, having preferences, representation, and intelligent problem-solving greatly reduces the toolkit of techniques we can use to understand and control its behavior. Consider the task of getting a pigeon to correctly distinguish videos of dance vs. those of martial arts. If one approaches the system bottom-up, one has to implement ways to interface to individual neurons in the animal's brain to read the visual input, distinguish the videos correctly, and then control other neurons to force the behavior of walking up to a button and pressing it. This may someday be possible, but not in our lifetimes. In contrast, one can simply train the pigeon [104]. Humanity has been training animals for millennia, without knowing anything about what is in their heads or how brains work. This highly efficient trick works because we correctly identified these systems as learning agents, which allows us to offload a lot of the computational complexity of any task onto the system itself, without micromanaging it.

What other systems might this remarkably powerful strategy apply to? For example, gene regulatory networks (GRNs) are a paradigmatic example of "genetic mechanism", often assumed to be tractable only by hardware (requiring gene therapy approaches to alter promoter sequences that control network connectivity, or adding/removing gene nodes). However, being open to the possibility that GRNs might actually be on a different place on this continuum suggests an experiment in which they are trained for new behaviors with specific combinations of stimuli (experiences). Indeed, recent analyses of biological GRN models reveal that they exhibit associative and several other kinds of learning capacity, as well as pattern completion and generalization [105; 106; 107; 108]. This is an example in which an empirical approach to the correct level of agency for even simple systems not usually thought of as cognitive suggests new hypotheses which in turn open a path to new practical applications (biomedical strategies using associative regimes of drug pulsing to exploit memory and address pharmacoresistance by abrogating habituation etc.).

We next consider specific aspects of the framework, before diving into specific examples in which it drives novel empirical work.

*Specific conceptual components of the TAME framework*

A useful framework in this emerging field should not only serve as a lens with which to view data and concepts [109], but also should drive research in several ways. It needs to first specify definitions for key terms such as a Self. These are not meant to be exclusively correct – the definitions can co-exist with others, but should identify a claim as to what is an essential invariant for Selves (and what other aspects can diverge), and how it intersects with experiment. The fundamental symmetry unifying all possible Selves should also facilitate direct comparison or even classification of truly diverse intelligences, sketching the markers of Selfhood and the topology of the option space within which possible agents exist. The framework should also help scientists derive testable claims about how borders of a given Self are determined, and how it interacts with the outside world (and other agents). Finally, the framework should provide actionable, semi-quantitative definitions that have strong implications and constrain theories about how Selves arise and change. All of this must facilitate experimental approaches to determine the empirical utility of this approach.

The TAME framework takes the following as the basic hallmarks of being a Self: the ability to pursue goals, to own compound memories, and to serve as the center of gravity for credit assignment (be rewarded or punished), where all of these are at a scale larger than possible for any of its components alone. Given the gradualist nature of the framework, the key question for any agent is "how well", "how much", and "what kind" of capacity it has for each of those key aspects, which in turn allows agents to be directly compared in an option space. TAME emphasizes defining a higher scale at which the (possibly competent) activity of component parts gives rise to an emergent system. Like a valid mathematical theorem which has a unique structure and existence over and above any of its individual statements, a Self can own for example associative memories (that bind into new mental content experiences that occurred separately to its individual parts), be the subject of reward or punishment for complex states (as a consequence of highly diverse actions that its parts have taken), and be stressed by states of affairs (deviations from goals or setpoints) that are not definable at the level of any of its parts (which of course may have their own distinct types of stresses and goals). These are practical aspects that suggest ways to recognize, create, and modify Selves.

Selves can be classified and compared with respect to the scale of goals they can pursue (Figure 4, described in detail in [110]). In this context, the goal-directed perspective adopted here builds on the work of Nagel, Mayr, and Rosenblueth et al. [4; 111; 112], emphasizing plasticity (ability to reach a goal state from different starting points) and persistence (tendency to reach that goal [113] state despite perturbations).

The ability of a system to expend energy to work toward a state of affairs, overcoming obstacles (to the degree that its sophistication allows) to achieve a particular set of substates is very useful for defining Selves because it grounds the question in well-established control theory and cybernetics (i.e., systems "trying to do things" is no longer magical but is well-established in engineering), and provides a natural way of discovering, defining, and altering the preferences of a system. A common objection is: "surely we can't say that thermostats have goals and preferences?". The TAME framework holds that <u>whatever true goals and preferences are, there must exist primitive, minimal versions from which they evolved</u>; simple homeostatic circuits are an ideal candidate for the "hydrogen atom" of goal-directed activity [4; 114]. It is logically inevitable that if one follows a complex cognitive capacity backwards through phylogeny, one eventually reaches precursor versions of that capacity that naturally suggest the (misguided) question "now is that *really* cognitive, or just physics?". Indeed a kind of goal-directedness permeates all of physics [115; 116; 117; 118; 119; 120], supporting a continuous climb of the scale and sophistication of goals.

Pursuit of goals is central to composite agency and the "many to one" problem because it requires distinct mechanisms (for measurement of states, storing setpoints, and driving activity to minimize the delta between the former and the latter) to be bound together into a functional unit that is greater than its parts. To co-opt a great quote [121], nothing in biology makes sense except in light of teleonomy [101; 111; 112; 113; 122; 123; 124; 125].

The expenditure of energy *in ways that effectively reach specific states despite uncertainty, limitations of capability, and meddling from outside forces* is proposed as a central unifying invariant for all Selves – a basis for the space of possible agents. This view suggests a semi-quantitative multi-axis option space that enables direct comparison of diverse intelligences of all sorts of material implementation and origins [12; 110]. Specifically (Figure 4), a "space-time" diagram can be created where the spatio-temporal *scale* of any agent's goals delineates that Self and its cognitive boundaries.

Note that the distances on Figure 4E represent not first-order capacities such as sensory perception (how far away can it sense), but second-order capacities of the size of goals (humble metabolic hunger-satiety loops or grandiose planetary-scale engineering ambitions) which a given cognitive system is capable of representing and working toward. At any given time, an Agent is represented by a single shape in this space, corresponding to the size and complexity of their possible goal domain. However, genomes (or engineering design specs) map to an ensemble of such shapes in this space because the boarders between Self and world, and the scope of goals an agent's cognitive apparatus can handle, can shift during the lifetime of some agents – "in software" (another "great transition" marker). All regions in this space can potentially define some possible agent. Of course, additional subdivisions (dimensions) can easily be added, such as the Unlimited Associative Learning marker [126] or aspects of Active Inference [127; 128; 129; 130].

Some agents, like a bacterium, have minimal memory and can concern themselves only with a very short time horizon and spatial radius – follow local gradients. Some agents, e.g., a dog, have more memory and more forward planning ability, but are still precluded from, for example, caring about what will happen 2 weeks hence, in the next town over. Some, like human beings, can devote their lives to causes of enormous scale. Akin to Special Relativity, this formalization makes explicit capacities that are forever inaccessible to a given agent (demarcating the edge of the "light cone" of its cognition).

In general, larger selves 1) are capable of working toward states of affairs that occur farther into the future (perhaps outlasting the lifetime of the agent itself – an important great transition, in the sense of [131], along the cognitive continuum); 2) deploy memories further back in time (their actions become less "mechanism" and more *decision-making* [132] because it's linked to a network of functional causes and information with larger diameter); and 3) they expend effort to manage sensing/effector activity in larger spaces (from subcellular networks to the extended mind [133; 134]). Overall, increases of agency is driven by mechanisms that scale up stress (Box 1) – the scope of states that an agent can possibly be stressed about (in the sense of pressure to take corrective action). In this framework, stress (as a system-level response to distance from setpoint states), preferences, motivation, and the ability to functionally care about what happens are tightly linked. Homeostasis, necessary for life, evolves into allostasis [135; 136] as new architectures allow tight, local homeostatic loops to be scaled up to measure, cause, and remember larger and more complex states of affairs [137].

Additional implications of this view are that Selves: are malleable (the borders and scale of any Self can change over time), can be created by design or by evolution, and are multi-scale entities that consist of other, smaller Selves. Indeed they are a patchwork of agents (akin to Theophile Bordeu's "many little lives" [138; 139]) that overlap with each other, and compete, communicate, and cooperate both horizontally (at their own level of organization) and vertically (with their component subunits and the super-Selves of which they are a part [140]).

Another important invariant for comparing diverse intelligences is that they are all solving problems, in some space (Figure 5). It is proposed that the traditional problem-solving behavior we see in standard animals in 3D space is just a variant of evolutionarily more ancient capacity to solve problems in metabolic, physiological, transcriptional, and morphogenetic spaces (as one possible sequential timeline along which evolution pivoted some of the same strategies to solve problems in new spaces). For example, when planaria are exposed to barium, a nonspecific potassium channel blocker, their heads explode. Remarkably, they soon regenerate heads that are completely insensitive to barium [141]. Transcriptomic analysis revealed that relatively few genes out of the entire genome were regulated to enable the cells to resolve this physiological stressor using transcriptional effectors to change how ions and neurotransmitters are handled by the cells. Barium is not something planaria ever encounter ecologically (so there should not be innate evolved responses to barium exposure), and cells don't turn over fast enough for a selection process (e.g., with bacterial persisters after antibiotic exposure). The task of determining which genes, out of the entire genome, can be transcriptionally regulated to return to an appropriate physiological regime is an example of an unconventional intelligence navigating a large-dimensional space to solve problems in real-time [142; 143; 144; 145]. Also interesting is that the actions taken in transcriptional space (a set of mRNA states) map onto a path in physiological state (the *ability* to perform many needed functions despite abrogated $K^+$ channel activity, not just a single state).

The common feature in all such instances is that the agent must navigate its space(s), preferentially occupying adaptive regions despite perturbations from the outside world (and from internal events) that tend to pull it into novel regions. Agents (and their sub- and super-agents) construct internal models of their spaces [53; 146; 147; 148; 149; 150; 151; 152; 153], which may or may not match the view of their action space developed by their conspecifics, parasites, and scientists. Thus, the space one is navigating is in an important sense virtual (belonging to some Agent's self-model), is developed and often modified "on the fly" (in addition to that hardwired by the structure of the agent), and not only faces outward to infer a useful structure of its option space but also faces inward to map its own body and somatotopic properties [154]. The lower-level subsystems simplify the search space for the higher-level agent because their modular competency means that the higher-level system doesn't need to manage all the microstates (a strong kind of hierarchical modularity [155; 156]). In turn, the higher-level system deforms the option space for the lower-level systems so that they do not need to be as clever, and can simply follow local energy gradients.

The degree of intelligence, or sophistication, of an agent in any space is then roughly proportional to its ability to deploy memory and prediction (information processing) in order to avoid local maxima. Intelligence involves being able to temporarily move away from a simple vector towards one's goals can often net bigger improvements down the line, but the agent's internal complexity has to facilitate some degree of complexity (akin to hidden layers in an ANN) in the goal-directed activity that enables the buffering needed for patience and indirect paths to the goal. This buffering enables the flip side of homeostatic problem-driven (stress reduction) behavior by cells: the exploration of the space for novel opportunities (creativity) by the collective agent, and the ability to acquire more complex goals (in effect, beginning the climb to Maslow's hierarchy [157]).

Another important aspect of intelligence that is space-agnostic is generalization. For example, in the barium planaria example discussed above, it is possible that part of the problem-solving capacity is due to the cells' ability to generalize depolarization induced by familiar epileptic triggers to be in the same class of physiological stresses as the totally novel barium effects, and deploy similar solutions. This has now been linked to measurement invariance [158], showing its ancient roots in the continuum of cognition.

Consistent with the above discussion, complex agents often consist of components that are themselves competent problem-solvers in their own (usually smaller, local) spaces. The relationship between wholes and their parts can be as follows. An agent is an integrated holobiont to the extent that it distorts the option space, and the geodesics through it, for its subunits (perhaps akin to how matter and space affect each other in general relativity) to get closer to a high-level goal in its space. A similar scheme is seen in neuroscience, where top-down feedback helps lower layer neurons to choose a response to local features by informing them about more global features [159].

At the level of the subunits, who know nothing of the higher problem space, this simply looks like they are minimizing free energy and passively doing the only thing they can do as physical systems: this is why if one zooms in far enough on any act of decision-making, all one ever sees is dumb mechanism and "just physics". The agential perspective [70] looks different at different scales of observation (and its degree is in the eye of a beholder who seeks to control and predict the system, which includes the Agent itself, and its various partitions). This view is closely aligned with that of "upper directedness" [160], in which the larger system directs its components' behavior by constraints and rewards for coarse-grained outcomes, not microstates [160].

Note that these different competing and cooperating partitions are not just diverse components of the body (cells, microbiome, etc.) but also future and past versions of the Self. For example, one way to achieve the goal of a healthier metabolism is to lock the refrigerator at night and put the keys somewhere that your midnight self, which has a shorter cognitive boundary (is willing to trade long-term health for satiety right now) and less patience, is too lazy to find. Changing the option space, energy barriers, and reward gradients for your future self is a useful strategy for reaching complex goals despite the shorter horizons of the other intelligences that constitute your affordances in action space.

The most effective collective intelligences operate by simultaneously distorting the space to make it easy for their subunits to do the right thing with no comprehension of the larger-scale goals, but themselves benefit from the competency of the subunits which can often get their local job done even if the space is not perfectly shaped (because they themselves are homeostatic agents in their own space). Thus, instances of communication and control between agents (at the same or different levels) are thus mappings between different spaces. This suggests that both evolution's, and engineers', hard work is to optimize the appropriate functional mapping toward robustness and adaptive function.

Next, we consider a practical example of the application of this framework to an unconventional example of cognition and flexible problem-solving: morphogenesis, which naturally leads to specific hypotheses of the origin of larger biological Selves (scaling) and its testable empirical (biomedical) predictions. This is followed with an exploration of the implications of these concepts for evolution, and a few remarks on consciousness.

**Somatic cognition: an example of unconventional agency in detail**

> "Again and again terms have been used which point not to physical but to psychical analogies. It was meant to be more than a poetical metaphor…"
> -- Hans Spemann, 1967

An example of TAME applied to basal cognition in an unconventional substrate is that of morphogenesis, in which the mechanisms of cognitive binding between subunits are now partially known, and testable hypotheses about cognitive scaling can be formulated (explored in detail in [99; 161; 162]). It is uncontroversial that morphogenesis is the result of collective activity: individual cells work together to build very complex structures. Most modern workers treat it as clockwork (with a few notable exceptions around the recent data on cell learning [163; 164; 165; 166; 167; 168; 169; 170]), preferring a purely feed-forward approach founded on the idea of complexity science and emergence. On this view, there is a privileged level of causation – that of biochemistry – and all of the outcomes are to be seen as the emergent consequences of highly parallel execution of local rules (a cellular automaton in every sense of the term). Of course, it should be noted that the forefathers of developmental biology, such as Spemann [171], were already well-aware of the possible role of cognitive concepts in this arena and others have occasionally pointed out detailed homologies [99; 172]. This becomes clearer when we step away from the typical examples seen in developmental biology textbooks and look at some phenomena that, despite the recent progress in molecular genetics, remain important knowledge gaps (Figure 6).

*Goal-directed activity in morphogenesis*

Morphogenesis (broadly defined) is not only a process that produces the same robust outcome from the same starting condition (development from a fertilized egg). In animals such as salamanders, cells will also *re*-build complex structures such as limbs, no matter where along the limb axis they are amputated, and *stop when it is complete*. While this regenerative capacity is not limitless, the basic observation is that the cells cooperate toward a specific, invariant endstate (the target morphology), from diverse starting conditions, and cease their activity when the correct pattern has been achieved. Thus, the cells do not merely perform a rote set of steps toward an emergent outcome, but modify their activity in a context-dependent manner to achieve a specific anatomical target morphology. In this, morphogenetic systems meet James' test for minimal mentality: "fixed ends with varying means" [173].

For example, tadpoles turn into frogs by rearranging their craniofacial structures: the eyes, nostrils, and jaws move as needed to turn a tadpole face into a frog face (Figure 6B). Guided by the hypothesis that this was not a hardwired but an intelligent process that could reach its goal despite novel challenges, we made tadpoles in which these organs were in the wrong positions – so-called Picasso Tadpoles [174]. Amazingly, they tend to turn into largely normal frogs because the craniofacial organs move in novel, abnormal paths (sometimes overshooting and needing to return a bit [175]) and stop *when they get to the correct frog face positions*. Similarly, frog legs that are artificially induced to regenerate create a correct final form but not via the normal developmental steps [176]. Students who encounter such phenomena and have not yet been inoculated with the belief that molecular biology is a privileged level of explanation [102] ask the obvious (and proper) question: how does it know what a correct face or leg shape is?

Examples of remodeling, regulative development (e.g., embryos that can be cut in half and produce normal monozygotic twins), and regeneration, ideally illustrate the goal-directed nature of cellular collectives. They pursue specific anatomical states that are much larger than any individual cells and solve problems in morphospace in a context-sensitive manner – any swarm of miniature robots that could do this would be called a triumph of collective intelligence in the engineering field. Guided by the TAME framework, two questions come within reach. First, how does the collective measure current state and store the information about the correct target morphology? Second, if morphogenesis is not at the clockwork level on the continuum of persuadability but perhaps at that of the thermostat, could it be possible to re-write the setpoint without rewiring the machine (i.e., in the context of a wild-type genome)?

*Pattern Memory: a key component of homeostatic loops*

Deer farmers have long known of trophic memory: wounds made on a branched antler structure in one year, will result in ectopic tines growing *at that same location* in subsequent years, long after the original rack of antlers has fallen off [177; 178; 179]. This process requires cells at the growth plate in the scalp to sense, and remember for months, the location of a transient damage event within a stereotypical branched structure, and reproduce it in subsequent years by over-riding the wild-type stereotypical growth patterns of cells, instead guiding them to a novel outcome. This is an example of experience-dependent, re-writable pattern memory, in which the target morphology (the setpoint for anatomical homeostasis) is re-written within standard hardware.

Planarian flatworms can be cut into multiple pieces, and each fragment regenerates precisely what is missing at each location (and re-scales the remaining tissue as needed) to make a perfect little worm [180]. Some species of planaria have an incredibly messy genome - they are mixoploid due to their method of reproduction: fission and regeneration, which propagates any mutations that don't kill the stem cell and expands it throughout the lineage (reviewed in [181]). Despite the divergence of genomic information, the worms are champion regenerators, with near 100% fidelity of anatomical structure. Recent data have identified one set of mechanisms mediating the ability of the cells to make, for example, the correct number of heads: a standing bioelectrical distribution across the tissue, generated by ion channels and propagated by electrical synapses known as gap junctions (Figure 7C,D). Manipulation of this voltage pattern by targeting the gap junctions or ion channels can give rise to planaria with one, two, or 0 heads, or heads with shape (and brain shape) resembling other extant species of planaria [182; 183]. Remarkably, the worms with abnormal head number are *permanently* altered to this pattern, despite their wild-type genetics: cut into pieces with no further manipulations, the pieces continue to regenerate with abnormal head number [41; 184]. Thus, much like the optogenetic techniques used to incept false behavioral memories into brains [185], modulation of transient bioelectric state is a conserved mechanism by which false pattern memories can be re-written into the genetically-specified electrical circuits of a living animal.

*Multi-scale competency of growth and form*
A key feature of morphogenesis is that diverse underlying molecular mechanisms can be deployed to reach the same large-scale goal. This plasticity and coarse-graining over subunits' states is a hallmark of collective cognition, and is also well known in neuroscience [186; 187]. Newt kidney tubules normally have a lumen of a specific size and are made up (in cross section) of 8-10 cells [188; 189]. When the cell size is experimentally enlarged, the same tubules are made of a smaller number of the bigger cells. Even more remarkable than the scaling of the cell number to unexpected size changes (on an ontogenetic, not evolutionary, timescale) is the fact that if the cells are made really huge, *just one cell* wraps around itself and still makes a proper lumen (Figure 6D). Instead of the typical cell-cell interactions that coordinate tubule formation, cytoskeletal deformations within one cell can be deployed to achieve the same end result. As in the brain, the levels of organization exhibit significant autonomy in the details of their molecular activity but are harnessed toward an invariant system-level outcome

*Specific parallels between morphogenesis and basal cognition*
The plasticity of morphogenesis is significantly isomorphic to that of brains and behavior because the communication dynamics that scale individual neural cells into a coherent Self are ones that evolution honed long before brains appeared, in the context of morphogenic control [181], and before that, in metabolic control in bacterial biofilms [190; 191; 192; 193]. Each genome gives rise to cellular hardware that implements signaling circuits with a robust, reliable default "inborn" morphology – just like genomes give rise to brain circuits that drive instinctual behavior in species that can build nests and do other complex things with no training.  However, evolution selected for hardware that can be reprogrammed by experiences, in addition to its robust default functional modes - in body structure, as well as in brain-driven behavior. Many of the brain's special features are to be found, unsurprisingly, in other forms outside the central nervous system. For example, mirror neurons and somatotopic representation are seen in limbs' response to injury, where the type and site of damage to one limb can be read out within 30 seconds from imaging the opposite, un-injured limbs [194]. Table 2 shows the many parallels between morphogenetic and cognitive systems.

*Not just philosophy: why these parallels matter*

The view of anatomical homeostasis as a collective intelligence is not a neutral philosophical viewpoint – it makes strong predictions, some of which have already borne fruit. It led to the discovery of reprogrammable head number in planaria [195] and of pre-neural roles for serotonin [196; 197]. It explains the teratogenicity for pre-neural exposure to ion channel or neurotransmitter drugs [198], the patterning defects observed in human channelopathies in addition to the neurological phenotypes (reviewed in [199]), and the utility of gap junction blockers as general anesthetics.

Prediction derived from the conservation and scaling hypotheses of TAME can be tested via bioinformatics. Significant and specific overlap are predicted for genes involved in morphogenesis and cognition (categories of memory and learning). This is already known for ion channels, connexin (GJ) genes, and neurotransmitter machinery, but TAME predicts a widespread re-use of the same molecular machinery. Cell-cell communication and cellular stress pathways should be involved in social behavior, while memory genes should be identified in genetic investigations of cancer, regeneration, and embryogenesis.

Another key prediction that remains to be tested (ongoing in our lab) is trainability of morphogenesis. The collective intelligence of tissues could be sophisticated enough to be trainable via reinforcement learning for specific morphological outcomes. Learning has been suggested by clinical data in the heart [200], bone [201; 202], and pancreas [203]. It is predicted that using rewards and punishments (with nutrients/endorphins and shock), not micromanagement of pathway hardware, could be a path to anatomical control in clinical settings, whether for morphology or for gene expression [106]. This would have massive implications for regenerative medicine, because the complexity barrier prevents advances such as genomic editing from impacting e.g, limb regeneration in the foreseeable future. The same reasons for which we would rather train a rat for a specific behavior than control all of the relevant neurons to force it to do it like a puppet, explain why the direct control of molecular hardware is a far more difficult biomedical path than understanding the sets of stimuli that could motivate tissues to build a specific desired structure.

The key lesson of computer science has been that even with hardware we understand (if we built it ourselves), it is much more efficient and powerful to understand the software and evince desired outcomes by the appropriate stimulation and signaling, not physical rewiring. If the hardware is reprogrammable (and it is here argued that much of the biological hardware meets this transition), one can offload much of the complexity onto the system itself, taking advantage of whatever competence the sub-modules have. Indeed, neuroscience itself may benefit from cracking a simpler version of the problem, in the sense of neural decoding, done first in non-neural tissues.

*Non-neural bioelectricity: what bodies think about*

The hardware of the brain consists of ion channels which set the cell's electrical state, and controllable synapses (e.g., gap junctions) which can propagate those states across the network. This machinery, including the neurotransmitters that eventually transduce these computations into transcriptional and other cell behaviors, is in fact highly conserved and present in all cells, from the time of fertilization (Figure 7C,D). A major difference between neural and non-neural bioelectricity is the time constant with which it acts (brains speed up the system into millisecond scales, while developmental voltage changes occur in minutes or hours [204; 205]). Key aspects of this system in any tissue that enable it to support flexible software is that both ion channels and gap junctions are themselves voltage sensitive – in effect, transistors (voltage-gated current conductances). This enables evolution to exploit the laws of physics to rapidly generate very complex circuits with positive (memory) and negative (robustness) feedback [206; 207; 208; 209; 210]. The fact that a transient voltage state passing through a cell can set off a cycle of progressive depolarization (like an action potential) or GJ closure means that such circuits readily form dynamical systems memories which can store different information and change their computational behavior without changing the hardware (i.e., not requiring new channels or gap junctions) [211], as is obvious in the action potential propagations in neural networks but is rarely thought about in development.

Consistent with its proposed role, slowly-changing resting potentials serve as instructive patterns guiding embryogenesis, regeneration, and cancer suppression [212; 213; 214]. In addition to the pattern memories encoded electrically in planaria (discussed above), they have also been shown to dictate the morphogenesis of the face, limbs, and brain, and function in determining primary body axes, size, and organ identity (reviewed in [215]). One of the most interesting aspects of developmental bioelectricity is its modular nature: very simple voltage states trigger complex, downstream patterning cascades. As in the brain, modularity goes hand-in-hand with pattern completion: the ability of such networks to provide entire behaviors from partial inputs. For example, Figure 7F shows how a few cells transduced with an ion channel that sets them into a "make the eye here" trigger recruit their neighbors, in any region of the body, to fulfill the purpose of the subroutine call and create an eye. Such modularity makes it very easy for evolution to develop novel patterns by re-using powerful triggers. Moreover, as do brains, tissues use bioelectric circuits to implement pattern memories that set the target morphology for anatomical homeostasis (as seen in the planarian examples above). This reveals the non-neural material substrate that stores the information in cellular collectives, which is a distributed, dynamic, re-writable form of storage that parallel recent discoveries of how group knowledge is stored in larger-scale agents such as animal swarms [216; 217]. Finally, bioelectric domains [218; 219; 220; 221; 222] set the borders for groups of cells that are going to complete a specific morphogenetic outcome – a system-level process like "make an eye". They define the spatio-temporal borders of the modular activity, and suggest a powerful model for how Selves scale in general.

*A bioelectric model of the scaling of the Self*

Gap junctional connections between cells provide an interesting case study for how the borders of the Self can expand or contract, in the case of a morphogenetic collective intelligence (Figure 8). Crucially, gap junctions (and gap junctions extended by tunneling nanotubes [223; 224]) enable a kind of cellular parabiosis – a regulated fusion between cells that enables lateral inheritance of physiological information, which speeds up processing in the same way that lateral gene inheritance potentiates change on evolutionary timescales. The following is a case study hypothesizing one way in which evolution solves the many-into-one problem (how competent smaller Selves bind into an emergent higher Self), and how this process can break down leading to a reversal (shrinking) of the Self boundary (summarized in Table 3).

Single cells (e.g., the protozoan *Lacrymaria olor*) are very competent in handling morphological, physiological, and behavioral goals on the scale of one cell. When connected to each other via gap junctions, as in metazoan embryos, several things happen (much of which is familiar to neuroscientists and workers in machine learning in terms of the benefits of neural networks) which lead to the creation of a Self with a new, larger cognitive boundary. First, when cells join into an electrochemical network, they can now sense events, and act, on a much larger physical "radius of concern" than a single cell. Moreover, the network can now integrate information coming from spatially disparate regions in complex ways that result in activity in other spatial regions. Second, the network has much more computational power than any of its individual cells (nodes), providing an IQ boost for the newly formed Self. In such networks, Hebbian dynamics on the electrical synapse (GJ) can provide association between action in one location and reward in another, which enables the system to support credit assignment at the level of the larger individual.

The third consequence of GJ connectivity is the partial dissolution of informational boundaries between the subunits. GJ-mediated signals are unique because they give each cell immediate access to the internal milieu of other cells. A conventional secreted biochemical signal arrives from the outside, and when it triggers cell receptors on the surface, the cell clearly knows that this information originated externally (and can be attended to, ignored, etc.) – it is easy to maintain boundary between Self and world. However, imagine a signal like a calcium spike originating in a cell due to some damage stimulus for example. When that calcium propagates onto the GJ-coupled neighbor, there are no metadata on that signal marking its origin; recipient cell only knows that a calcium spike occurred, and cannot tell that this information does not belong to it. The calcium spike (and its downstream effects) is a kind of false memory for the recipient cell, but a true memory for the collective network of the stimulus that occurred in one part of the individual. This wiping of ownership information for GJ signals as they propagate through the network is critical to enabling a partial "mind meld" between the cells: keeping identity (in terms of distinct individual history of physiological states – memory) becomes very difficult, as small informational molecules propagate and mix within the network. Thus, this property of GJ coupling promotes the creation of a larger Self by partially erasing the mnemic boundaries between the parts which might impair their ability to work toward a common goal. This is a key part of the scaling of the Self by enlarging toward common goals – not by micromanagement, but by bringing multiple subunits into the same goal-directed loop by tightly coupling the sensing, memory, and action steps in a syncytium where all activity is bound toward a system-level teleonomic process. When individual identities are blurred in favor of longer time-scale, larger computations in tissues, small-horizon (myopic) action in individual cells (e.g., cancer cells' temporary gains followed by maladaptive death of the host) leads to a more adaptive longer-term future as a healthy organism. In effect, this builds up long-term collective rationality from the action of short-sighted irrational agents [225; 226].

It is important to note that part of this story has already been empirically tested in assays that reveal the shrinking as well as the expansion of the Self boundary (Figure 9). One implication of these hypotheses is that the binding process can break down. Indeed this occurs in cancer, where oncogene expression and carcinogen exposure leads to a closure of GJs [227; 228]. The consequence of this is transformation to cancer, where cells revert to their ancient unicellular selves [229] – shrinking their computational boundaries and treating the rest of the body as external environment. The cells migrate at will and proliferate as much as they can, fulfilling their cell-level goals – metastasis (but also sometimes attempting to, poorly, reboot their multicellularity and make tumors [230]). The model implies that this phenotype can be reverted by artificially managing the bioelectric connections between a cell and its neighbors. Indeed, recent data show that managing this connectivity can override default genetically-determined states, inducing metastatic melanoma in a perfectly wild-type background [231] or suppressing tumorigenesis induced by strong oncogenes like p53 or KRAS mutations [232]. The focus on physiological connectivity (information dynamics) – the software – is consistent with the observed facts that genetic alterations (hardware) are not necessary to either induce or revert cancer (reviewed in [233]).

All these dynamics lead to a few interesting consequences. GJ-mediated communications are not merely conversations (in the way that external signaling is) – they are binding, in the sense that once a GJ is open, a cell is subject to whatever comes in from the neighbor. In the same sense, having a synapse makes one vulnerable to the state of neighbors. GJs spread (dilute) the pain of depolarization, but at the same time give a cell's neighbors the power to change its state. Compatible with the proposal that the magnitude of a Self is the scale and complexity of states it can be stressed by, connections by tunable, dynamic GJs greatly expand the spatial, temporal, and complexity of things that can irritate cells – complex events from far away can now percolate into a cell via non-linear GJ paths through the network, enabling the drive to minimize such events now necessarily involves homeostatic activity of goal states, sensing, and activity on a much larger scale. Stress signals, propagating through such networks, incentivize other regions of the tissue to act cooperatively in response to *distant* events by harnessing their selfish drive to reduce their own stress. This facilitates the coherent, system-level response to perturbations beyond their local consequences, and gives rise to larger Selves able to react coherently to stressful departures from more complex, spatially-distributed allostatic setpoints. For example, whereas a solitary cell might be stressed (and react to) abnormal local pH, cells that are part of a transplanted salamander limb will be induced to a more grandiose activity: change the number of fingers they produce to be correct relative to the limb's new position in the host's body [234], a decision that involves large-scale sensing, decision-making, and control.

A fourth consequence of the coupling is that cooperation in general is greatly enhanced. In the game theory sense, it is impossible to cheat against your neighbor if you are physiologically coupled. Any positive or negative effects of a cell's actions toward the neighboring cell are immediately propagated back to it, in effect producing one individual in which the parts cannot "defect" against each other. This dynamic suggests an interesting twist on Prisoners' Dilemma models in which the number of agents is not fixed, because they have the options of Cooperate, Defect, Merge, and Split (we are currently analyzing such models). Specifically, merging with another agent creates an important dimensionality reduction (because defection is no longer an option); this not only changes the calculus of game theory as applied to biological interactions, but also the action space itself. These dynamics take place on a developmental timescale, complementing the rich existing literature on game theory in evolution [235; 236; 237; 238].

Indeed, the smaller and larger agents' traversal of their various spaces provides a way to think about how smaller agents' (cell-level) simple homeostatic loops can scale up into large, organ-level anatomical homeostatic loops. Prentner recently showed how agents build up spatial models of their worlds by taking actions that nullify changes in their experience [151]. Nullifying changes is the core of homeostasis, which together with the closely related surprise minimization framework [239; 240; 241], suggests a straightforward sense in which larger Selves scale up to models of their world and themselves from evolutionary primitives such as metabolic homeostasis. Bioelectricity provides examples where cell-level physiological homeostats form networks that implement much larger-scale pattern memories as attractors, akin to Hopfield networks (Figure 10) [207; 208; 209; 210; 211; 242; 243]. This enables all tissues to participate in the kind of pattern completion seen in neural networks – a critical capability for regenerative and developmental repair (anatomical homeostasis).

With these pieces in place, it is now possible to visualize the progressive scaling that expands the cognitive light cone. Cells with a chemical receptor can engage in predictive coding to manage their sensory experience [239; 244; 245]. Taking one step further, individual cells homeostatically maintaining $V_{mem}$ (cell membrane resting potential voltage) levels readily create bioelectric networks that work as a kind of virtual governor, maintaining patterns of $V_{mem}$ (which is a coarse-grained parameter, over many different details of ion concentrations and ion channel states) against perturbation with greater stability [207; 208; 209; 210; 211; 220; 246; 247]. Moreover, responding to specific levels of cellular $V_{mem}$ begins the journey to generalization – using built-in responses in light of novel combinations of stimuli (e.g., familiar depolarization events caused by novel ion dynamics). Gap junctions propagate voltage states across tissue, allowing cells to respond to events that are not local in nature (larger-scale) and to respond *en masse*. More generally, this means that the input to any group of cells is produced by the output of groups of cells – sub-networks, which can be complex and highly processed over time (not instantaneous), enabling predictive coding to manage complex states (at a distance in space and time) and not only purely local, immediate sensory data. It also means that the system is extremely modular, readily connecting diverse upstream and downstream events to each other via the universal adapter of bioelectric states. When this is applied to the homeostatic TOTE (test-operate-exit) loop, allowing its measurement, comparison, and action modules to be independently scaled up (across space, time, and complexity metrics), this inherently expands the cognitive light cone of a homeostatic agent to enable progressively more grandiose goals.

Crucially, all of the above-mentioned aspects of the role of generic bioelectric networks underlying the scaling of Selves are not only the products of the evolutionary process, but have many functional implications for evolution itself (forming a positive feedback loop in which rising multiscale agency potentiates the evolution of increasingly more complex versions).

**Evolutionary aspects**

Developmental bioelectricity works alongside other modalities such as gene-regulatory networks, biomechanics, and biochemical systems. The TAME framework emphasizes that what makes it special is that it's not just another micro-mechanism that developmental biologists need to track. First, developmental bioelectrics is a unique computational layer that provides a tractable entrypoint into the informational architecture and content of the collective intelligence of morphogenesis. Second, bioelectric circuits show examples of modularity, memory, spatial integration, and generalization (abstraction over ion channel microstates) – critical aspects of understanding basal origins of cognition Developmental bioelectricity provides a bridge between the early problem-solving of body anatomy and the more recent complexity of behavioral sophistication via brains. This unification of two disciplines, suggests a number of hypotheses about the evolutionary path that pivoted morphogenetic control mechanisms into the cognitive capacities of behavior, and thus sheds light on how Selves arise and expand.

*Somatic bioelectrics reveals the origin of complex cognitive systems*

Developmental bioelectrics is an ancient precursor to nervous systems. Analog bioelectrical dynamics generate patterns in homogenous cell sheets and coordinate information that regulates transcription and cell behaviors. Evolution first exploited this to enable cell groups to position the body configuration in developmental morphospace, long before some cells specialized to use very fast, digital spiking as neural networks for control of behavior as movement in 3-dimensional space [181]. The function of nervous systems as spatial organizers operating on data from the external world [248] is an adaptation built upon the prior activity of bioelectric circuits in organizing the internal morphology by processing data from the internal milieu. While neural tissues electrically encode spatial information to guide movement (e.g., memory of a maze in a rat brain) by controlling muscles, bioelectric prepatterns guide the behaviors of other cell types, on slower timescales, during development, regeneration, and remodeling toward invariant, robust anatomical configurations.

Developmental bioelectricity illustrates clearly the continuous nature of properties thought to be important for cognition, and the lack of a clean line separating brainy creatures from others. On a single-cell level, even defining a "neuron" is not trivial, as most cells possess the bioelectrical machinery and a large percentage of neuronal genes are also expressed in non-neural cells [249], while neural molecular components are found in cytonemes [250]. Many channel families were likely already present in the most recent unicellular ancestor [84]. The phylogeny of ion channels is ancient, and the appearance of context-sensitive channels (enabling new kinds of bioelectrical feedback loops) tracks well with the appearance of complex body plans at the emergence of metazoa [83], revealing the remarkable evolutionary continuum that leads from membrane excitability in single cells to cognitive functions in advanced organisms, by way of somatic pattern control [86].

Fascinating work on bacteria has shown that prokaryotes also utilize bioelectric state for proliferation control [251]; and, paralleling the developmental data discussed above, bioelectric phenomena in bacteria scale easily from single-cell properties [252] to the emergence of proto-bodies as bacterial biofilms. Bacterial communities use brain-like bioelectric dynamics to organize tissue-level distribution of metabolites and $2^{nd}$ messenger molecules, and illustrating many of the phenomena observed in complex morphogenetic contexts, such as encoding stable information in membrane potential patterns, bistability, and spatial integration [190; 192; 253; 254; 255]. Not only animal lineages, but plants [256; 257] [258] use bioelectricity, as evolution frequently exploits the fact that bioelectric dynamics are a powerful and convenient medium for the computations needed to solve problems in a variety of spaces not limited to movement in 3D space.

Developmental bioelectricity helps understand how free-living cells scaled their cell-level homeostatic pathways to whole body-level anatomical homeostasis [110]. It has long been appreciated that evolvability is potentiated by modularity – the ability to trigger complex morphogenetic cascades by a simple "master" trigger that can be re-deployed in various contexts in the body [259]. Recent advances reveal that bioelectric states can form very powerful master inducers that initiate self-limiting organogenesis. For example, the action of a single ion channel can induce an eye-specific bioelectric state that creates complete eyes in gut endoderm, spinal cord, and posterior tissues [260] – locations where genetic "master regulators" like the *Pax6* transcription factor are insufficient in vertebrates [261]. Likewise, misexpression of a proton pump (or a 1-hour ionophore soak) to trigger bioelectric changes in an amputation wound can induce an entire 8-day cascade of building a complete tadpole tail [262; 263]. This is control at the level of organ, not single cell fate specification, thus not requiring the experimenter to provide all of the information needed to build the complex appendage. Thus, bioelectric states serve as effective master regulators that evolution can exploit to make modular, large-scale changes in anatomy.

Moreover, because the same $V_{mem}$ dynamics can be produced by many different ion channel combinations, and because bioelectric states propagate their influence across tissue distance during morphogenesis [264; 265], evolution is free to swap out channels and explore the bioelectrical state space: simple mutations in electrogenic genes can exert very long-range, highly coordinated changes in anatomy. Indeed, the KCNH8 ion channel and a connexin were identified in the transcriptomic analysis of the evolutionary shift between two functionally different morphologies of fin structures in fish [266]. The evolutionary significance of bioelectric controls can also be seen across lineages, as some viruses evolved to carry ion channel and gap junction (Vinnexin) genes that enable them to hijack bioelectric machinery used by their target cells [267; 268].

The unique computational capabilities of bioelectric circuits likely enabled the evolution of nervous systems, as specialized adaptations of ancient ability of all cell networks to process electrical information as pre-neural networks [181; 269]. A full understanding of nervous system function must involve not only its genetics and molecular biology but also the higher levels of organization comprising dynamic physiology and computations involved in memory, decision-making, and spatio-temporal integration. The same is true for the rest of the body. For example the realization that epithelia are the generator of bioelectric information [270] suggests models in which they act like a retina wrapped around a whole embryo (and individual organs) to preprocess electrical signals into larger-scale features and compute contrast information for downstream processing [172]. The investigation of somatic bioelectric states as primitive "pattern memories" and the expansion of computational science beyond neurons will enrich the understanding of cell biology at multiple scales beyond molecular mechanisms, as is currently only done with respect to the brain [271]. Generalizing the deep concepts of multiscale neuroscience beyond neurons [99; 109; 172] is necessary for a better understanding of the tissue-level decision-making that drives adaptive development and regeneration. Conversely, advances in understanding information processing in a relatively simpler anatomical context will feed back to enrich important questions in brain science, shedding light on fundamental mechanisms by which information-processing agents (cells) work collectively to accomplish unified, complex system-level outcomes. The multi-disciplinary opportunity here is not only to gain insight into the phylogeny of nervous systems and behavior, but to *erase the artificial boundaries between scientific disciplines that focus on neurons vs. the rest of the body, with the direct consequence that a more inclusive, gradualist picture emerges of the mechanisms commonly associated with cognitive Selves*.

Ion channels and gap junctions are the hardware interface to the bioelectric computational layer within living systems. Like a retina for a brain, or a keyboard for a computer, they allow transient signals to serve as inputs to memory and decision-making networks. For any given agent (cell, tissue, etc.), its bioelectrical interface is accessed by a number of potential users. First are the neighboring agents, such as other tissues, which pass on their bioelectric state during cooperative and competitive interactions in morphogenesis. There are also commensal and parasitic microbes, which have evolved was to hijack such control systems to manipulate the anatomy of the host – like the naïve bacteria on planaria that can determine head number and visual system structure in flatworm regeneration [272]. Moreover, the development of pharmacological, genetic, and optogenetic tools now allows human bioengineers to access bioelectrical circuits for the control of growth and form in regenerative medicine and synthetic bioengineering contexts [273; 274; 275; 276; 277; 278]. All of these manipulations can serve as catalysts, enabling an evolutionary lineage to more easily travel to regions of option space that might otherwise be separated by an energy barrier that is difficult for standard evolution to reach. We next look at specific ways in which the architecture of multiscale autonomy, especially as implemented by bioelectric network mechanisms, potentiates evolution.

*Multi-scale autonomy potentiates the speed of evolution*

Deterministic chaos and complexity theory have made it very clear why bottom-up control of even simple systems (e.g., 3-body problem) can be practically impossible. This inverse problem [179] – what control signals would induce the desired change – is not only a problem for human engineers but also for adjacent biological systems such as the microbiome or a fungus that seeks to control the behavior of an ant [279], and most of all, for other parts of a complex system (to help control itself). Evolution tackles this task by exploiting a multiscale competency architecture (MCA), where subunits making up each level of organization are themselves homeostatic agents. It's built on an extremely powerful design principle: error correction [280; 281; 282].

The key aspect of biological modularity is not simply that complex subroutines can be triggered by simple signals, making it easy to recombine modules in novel ways [283;

284], but that these modules are also themselves sub-agents exhibiting infotaxis and socialtaxis, and solving problems in their own spaces [285; 286] [287]. When an eye primordium appears in the wrong place (e.g., a tadpole tail), it still forms a correctly patterned, functional organ, manages to get its data to the brain (via spinal cord connection) to enable vision (Fig. 6E), and (if somewhere in the head) moves to the correct place during metamorphosis [174]. When cells are artificially made to be very large and have several times the normal genetic material, morphogenesis adapts to this and still builds an overall correct animal [188; 189]. These are goal-directed (in the cybernetic sense) processes because the system can reach a specific target morphology (and functionality) state despite perturbations or changes in local/starting conditions or the basic underlying components. Regeneration is the most familiar example of this, but is just a special case of the broader phenomenon of anatomical homeostasis. This ability has several (closely related) key implications for the power and speed of evolution (summarized in Table 4).

First, it greatly smooths the fitness landscape. Consider two types of organisms: one whose subsystems mechanically follow a hardwired (genetically-specified) set of steps (A, passive, or merely structural modularity), and one whose modules optimize a reward function (B, multi-scale competency of modules). Mutations that would be detrimental in A (e.g., because they move the eye out of its optimal position) are neutral in B, because the competency of the morphogenetic subsystems repositions the eye even if it starts out somewhere else. Thus, MCA shields from selection some aspects of mutations' negative effects (which inevitably are the bulk of random mutations' consequences). The primary reason for the anatomical homeostasis seen in regulative development and regeneration may be not for dealing with damage, but deviations from target morphology induced by mutations. This is certainly true at the scale of tissues during the lifetime of an individual (as in the inverse relationship between regeneration and cancerous defection from large-scale target morphology [229]), but may be true on evolutionary time scales as well.

Second, MCA reduces apparent pleiotropy – the fact that most mutations have multiple effects [288]. For example, a change in an important signaling pathway such as Wnt or BMP is going to have numerous consequences for an organism. Suppose a mutation appears that moves the mouth off of its optimal position (bad for fitness) but also has some positive (adaptive) feature elsewhere in the body. In creatures of type A, the positive aspects of that mutation would never be seen by selection because the malfunctioning mouth would reduce the overall fitness or kill the individual outright. However, in creatures of type B, the mouth could move to its optimal spot [174], enabling selection to independently evaluate the other consequence of that mutation. Creatures possessing MCA could reap the benefit of positive consequences of a mutation while masking its other effects via local adjustments to new changes that reduce the penalties (an important kind of buffering). In effect, evolution doesn't have to solve the very difficult search problem of "how to improve feature X without touching features Y...Z which already work well", and reaps massive efficiency (time) savings by not having to wait until the search process stumbles onto a way to directly encode an improvement that is either isolated from other features, or improves them all simultaneously [289; 290].

Third, MCA allows systems to not only solve problems, but to also exploit opportunities. A lineage has the chance to find out what pro-adaptive things a mutation can do, because competency hides the negative consequences. This gives time for new mutations to appear that hardwire the compensatory changes that had to be applied - an exact analogy to the Baldwin effect, where evolution eventually hardwires useful behaviors that eventually arose by learning. This enables the opportunity to exploit the possibility space more freely, providing a kind of patience or reduction of the constraint that evolutionary benefits have to be immediate in order to propagate – it effectively reduces the short-sightedness of the evolutionary process. Indeed, multiscale competency is beneficial not only for natural evolution, but also for soft robotics and synthetic bioengineering because it helps cross the sim-to-real gap: models do not have to be 100% realistic to be predictively useful if the component modules can adaptively make up for some degree of deficiency the controller design [291].

Fourth, the homeostatic setpoint-seeking architecture makes the relationship between genotype and anatomical phenotype more linear [179] [292], improving controllability [293; 294; 295]. By using a top-down control layer to encode the patterns to which competent subunits operate, living systems do not need to solve the difficult inverse problems of what signals to send their subsystems to achieve high-level outcomes. Bioelectric pattern memories (such as the voltage distribution that tells wild-type planarian cells whether to build 1 head or 2) exploit a separation of data from the machine itself, which makes it much easier to make changes. Evolution does not need to find changes at the micro level but can also simply change the information encoded in the setpoints, such as the electric face prepattern [174], which allows it to re-use the same exact implementation machinery to build something that can be quite different. The ability to rely on a non-zero IQ for your component modules (thus delegating and offloading complex regulatory chains) is an important affordance [108; 296] for the evolutionary process. It means that the larger system's evolution is in effect searching an easier, less convoluted control, signaling, or reward space – this massive dimensionality reduction offers the same advantages human engineers have with agents on the right side of the persuadability scale. It is no accident that learning in the brain, and behavioral systems, eventually exapted this same architecture and indeed the exact same bioelectrical machinery to speed up the benefits of evolution.

A significant brake on the efficiency of evolution, as on machine learning (indeed, all learning) is credit assignment: which change or action led to the improvement or reward? When a collection of cells known as a 'rat' learns to press a level and get a reward, no individual cell has the experience of interacting with a lever and receiving the nutrient. What enables the associative memory in this collective intelligence are the delay lines (nervous system) between the paws and the intestinal lining which provide a kind of patience – a tolerance of the temporal delay between the action and the reward and the ability to link extremely diverse modules on both ends (different kinds of actions can be linked to arbitrary rewards). MCA does the same thing for evolutionary learning [105; 297; 298; 299; 300], making it easier for systems to reap selection rewards for arbitrary moves in genotype space. This effectively raises the IQ of the evolutionary search process. Much as (Figure 5) an agent's sophistication can be gauged by how expertly and efficiently it navigates an arbitrary search space and its local optima, the traversal of the evolutionary search process can be made less short-sided by homeostatic activity within the physiological layer that sits between genotype and phenotype.

There is an adaptation tradeoff between robustness (e.g., morphogenesis to the same pattern despite interventions and changing conditions) and responsiveness to environment (context sensitivity), perhaps similar to the notion of criticality [301; 302]. The plasticity and goal-directedness of modules (as opposed to hardwired patterns) serve to reduce the sim-to-real gap [303]: because the current environment always offers novel challenges compared to prior experiences which evolution (or human design) uses to prepare responses, the MCA architecture doesn't take history too seriously, relying on plasticity and problem-solving more than on fine-tuning micromodels of what to do in specific cases. Biology reaps the benefits of both types of strategies by implementing anatomical homeostasis that coarse-grains robustness by making stability applying to large outcomes, such as overall anatomy, not to the microdetails of cell states. The scaling of homeostatic loops makes it possible to achieve both: consistent results and environmental sensitivity. These dynamics apply in various degrees to the numerous nested, adjacent, and overlapping sub-agents that make up any biological system. Cooperation results not from altruistic actions between Selves, but by the expansion of the borders of a single Self via scaling of the homeostatic loops. On this view, cancer cells are not more selfish than tissues – they are all equally selfish, but maintain goals appropriate to smaller scales of Selves. Indeed, even the parts of one normal body don't perfectly cooperate – this is as true in development [304] as it is in cognitive science [305; 306]. A picture is emerging of how evolution exploits the local competency of modules, competing and cooperating, to scale these subsystems' sensing, actuation, and setpoint memories to give rise to coherent larger-scale Selves. Overall, the TAME framework addresses functional aspects only, and is compatible with several views on phenomenal consciousness in compound Selves [307]. However, it does have a few implications for the study of Consciousness.

**Consciousness**
The ancient question of "where does it all come together?" in the brain, with respect to the unified character of consciousness, is one of those pseudo-problems that is immediately dispelled by a framework like TAME that focuses on multi-scale architecture. How big should a place where it all comes together be? If it can be ~140 mm wide, then the answer is, the whole brain. One could decide that it should be smaller (the human pineal gland is ~7 mm wide), but then the question is, why not smaller still – given the cellular components of the pineal (or any piece of the brain), one would always need to ask "but where does it all come together *inside there*?" of whatever piece of the brain is taken to be the seat of consciousness. The multi-scale nature of biology means that there is no privileged size scale for any homunculus.

Another important idea with respect to consciousness is "What is it like to be" a given agent [308]. Sensory augmentation, neural link technologies, and bioengineering produce tractable model systems in novel cognitive architectures, such as 2-headed planaria where the brains are connected by a central nervous system (Figure 7B), to help study the functional aspects of this cognitive re-shuffling. TAME's focus on cognitive architectures all inevitably being composites, emphasizes that the parts can be rearranged and thus that the Subject of cognition can change "on the fly", not merely during evolutionary timescales. Thus, the basic question of philosophy of mind – what's it like to be animal X [308] – is just a first-order step on a much longer journey. The second-order question is, what's it like to be a caterpillar, slowly *changing* into a butterfly as its brain is largely dissolved and reassembled into a different architecture for an animal whose sense organs, effectors, and overall Umwelt is completely different. All of this raises fascinating issues of first person experience not only in purely biological metamorphoses (such as human patients undergoing stem cell implants into their brains), but also technological hybrids such as brains instrumentized with novel sensory arrays, robotic bodies, software information systems, or functionally linked to other brains.

Most surprisingly, the plasticity and capacity for bioengineering and chimerization (recombination of biological and engineered parts in novel configurations) erases the sharp divide between first person and third person perspectives. This has been a fundamental, discrete distinction ever since Descartes, but the capacity for understanding and creating new combinations shows a continuum even in this basic distinction (Figure 11). The fact that Selves are not monadic means we can share parts with our subject of inquiry. If one has to *be* a system in order to truly know what it's like to be that system ($1^{st}$ person perspective), this is now possible, to various degrees, by physically merging one's cognitive architecture with that of another system. Of course, by strongly coupling to another agent, one doesn't remain the same and experience the other's consciousness; instead, a new Self is created that is a composite of the two prior individuals and has composite cognition. This is why consciousness research is distinct in strong degree from other scientific topics. One can observe gauges and instruments for $3^{rd}$-person science and remain the same Self (largely; the results of the observation may introduce small alterations in the cognitive structure). However, data on $1^{st}$ person experiential consciousness cannot be taken in without fundamentally changing the Self (being an effective homunculus by watching the neuroscience data corresponding to the movies inside the heads of other people is impossible for the same reason that there is no homunculus in each of our heads). The study of consciousness, whether done via scientific tools or via the mind's own capacity to change itself, inevitably alters the Subject.

With respect to the question of *consciousness per se*, as opposed to neural or behavioral correlates of consciousness, we have one major functional tool: general anesthesia. It is remarkable that we can readily induce a state in which all the individual cells are fine and healthy, but the larger Self is simply gone (although, some of the parts can continue to learn during this time [309]). Interestingly, general anesthetics are gap junction blockers [310]: consistent with the cognitive scaling example above, shutting down electrical communication among the cells leads to a disappearance of the higher-level computational layer while the cellular network is disrupted. GJ blockers are used to anesthetize living beings ranging across plants, Hydra, and human subjects [311]. It is amazing that the same Self (with memories and other properties) returns, when the anesthetic is removed. Of course, the Self does not return immediately, as shown by the many hallucinatory [312; 313] experiences of people coming out of general anesthesia – it takes some time for the brain to return to the correct global bioelectric state once the network connections are allowed again (meta-stability) [314]. Interestingly, and in line with the proposed isomorphism between cognition and morphogenesis, gap junction blockade has exactly this effect in regeneration: planaria briefly treated with GJ blocker regenerate heads of other species, but eventually snap out of it and remodel back to their correct target morphology [183]. It is no accident that the same reagents cause drastic changes in the high-level Selves in both behavioral and morphogenetic contexts: evolution uses the same scheme (GJ-mediated bioelectrical networks) to implement both.

The epistemic problem of Other Minds has been framed to imply that we cannot directly ever be sure how much or what kind of consciousness exists in any particular system under study. The TAME framework reminds us that this is true even for components of ourselves (like the non-verbal brain hemisphere). Perhaps the confabulation system enables one part of our mind to estimate the agency of other parts (the feelings of consciousness and free will) and develop models useful for prediction and control, applying in effect the empirical criteria for persuadability internally. The ability to develop a "theory of mind" about external agents can readily be turned inward, in a composite Self.

Are all cognitive systems conscious? The TAME framework is compatible with several views on the nature of consciousness. However, the evolutionary conservation of mechanisms between brains and their non-neural precursors has an important consequence for the question of where consciousness could be found. To the extent that one believes that mechanisms in the brain enable consciousness, all of the same machinery and many similar functional aspects are found in many other places in the body and in other constructs. TAME emphasizes that there is no principled way to restrict consciousness to "human-like, full-blown sophisticated brains", which means one has to seriously entertain degrees of consciousness in other organs, tissues, and synthetic constructs that have the same features neurons and their networks do [90; 315; 316; 317]. The fundamental gradualism of this framework suggests that whatever consciousness is, some variant and degree thereof has to be present very widely across autopoietic systems. TAME is definitely incompatible with binary views that cut off consciousness at a particular sharp line and it suggests no obvious way to define cognitive systems that have no consciousness whatsoever. A big open question is whether the continuum of cognition (and consciousness) contains a true "0" or only infinitisimal levels for very modest agents. One is tempted to imagine what properties a truly minimal agent has to have; unpredictability and ability to pursue goals seem key, and both of these are present to a degree in even single particles (via quantum indeterminacy and least action behavior). The scaling (or lack thereof) of these capacities in bulk inorganic matter vs. highly-organized living forms is a fertile area for future development of TAME and will be explored in forthcoming work.

**Conclusion**

*A more inclusive framework for cognition*

Regenerating, physiological, and behaving systems use effort (energy) to achieve defined, adaptive outcomes despite novel circumstances and unpredictable perturbations. That is a key invariant for cognition; differences in substrate, scale, or origin story among living systems are not fundamental, and obscure an important way to unify key properties of life: the ability to deploy intelligence for problem-solving in diverse domains. Modern theories of Mind must eventually handle the entire option space for intelligent agents, which not only contains the familiar advanced animals we see on Earth, but can also subsume ones consisting of radically different materials, ones created by synthetic bioengineering or combinations of evolution and rational design in the lab, and ones of exobiological as well as possible terrestrial origins. The advances of engineering confirm and put into practice an idea that was already entailed by evolution: that cognitive traits, like all other traits, evolved from humbler variants, forming a continuum. There are no truly separated natural kinds in this space, no scientifically-valid binary categories in this space. Like the definition of human "adults", which snap into being at the age of 18, binary views on cognitive properties are fictitious coarse-grainings useful for our legal system to operate, but no more than that. There is no bright line between "truly cognitive" and "pseudo cognitive" that can ever be drawn between two successive members of an evolutionary lineage. "Anthropomorphism" is a pseudo-scientific "folk" notion to be replaced by a view more informed by evolution and engineering-based approaches that empirically ask what level of cognitive model enables the most fruitful prediction, control, and novel experiments.

Every intelligence is a collective intelligence, and the modular, multi-scale architecture of life means that we are a holobiont in more than just the sense of having a microbiome [318] – we are all patchworks of overlapping, nested, competing, and cooperating agents that have homeostatic (goal-directed) activity within their self-constructed virtual space at a scale that determines their cognitive sophistication. A highly tractable model system for unconventional cognition, in which these processes and the scaling of Selves can not only be seen but also manipulated, is morphogenetic homeostasis. The construction and remodeling toward anatomical features of cellular collectives has crucial isomorphism to cognitive aspects of the many-into-one binding like credit assignment, learning, stress reduction, etc. The partial wiping of ownership information on permeant signals makes gap junctional coupling an excellent "hydrogen atom" for thinking about biological mechanisms that scale cognition while enable co-existence of subunits with local goals (multiple levels of overlapping Selves, whose scale and borders are porous and can change during the lifetime of the agent). However, many other substrates can no doubt fulfill the same functions.

*Big questions: conceptual next steps*

The TAME framework is conceptually incomplete in important ways. On-going development is proceeding along lines including:
- Merging with other frameworks, to develop plausible accounts of which minimal capacities are sufficient for the evolution of which others (e.g., is surprise minimization or infotaxis the origin, or a consequence, of homeostatic scaling). Connections should be explored to such as Active Inference [120; 241; 319] and Rosen's (M,R) and Anticipatory Systems [320; 321; 322; 323], and to recent advances in information theory as applied to individuality and scaling of causal power [324; 325; 326; 327].
- Improvement of the Cognitive Light cone space, to merge it quantitatively with the models of the expansion of goal representations, credit assignment, and metacognition, and mapping the specific waypoints in the axis of persuadability to transitions of scale with respect to the cognitive boundary metric, perhaps linking with existing models of Major Transitions [235; 328; 329].
- Development of methods to quantify the navigation of various spaces by diverse agents, especially to characterize the degree of look-ahead, planning, avoidance of local minima, generalization, etc. This will lead to an integration with the Wiener and Rosenbleuth scale (Figure 1B). More generally a key future aspect to address is whether these spaces have any objective reality, or whether the spaces that scientists see systems solving, as well as the spaces that the system itself constructs from its experience [150; 151], are equally valid and all virtual. One possible approach is to construct unsupervised automated methods (machine learning) that observe systems and construct models of spaces that can be evaluated for predictive quality and parsimony (a sort of Principal Component Analysis).

- Extension of the problem-solving and stress-guided paradigm by incorporating a notion of exploration, not driven by attempts to overcome current challenges but rather by a capacity (and pressure?) to explore better possibilities (complementing the drive to solve problems).
- Connection with the work on the robustness paradox [281; 282; 330; 331; 332], tracking how noise and stochasticity expand at lower levels in the presence of MCA which can reduce the power of selection for precision; doing this at the transcriptional and bioelectric network levels to understand how their degrees of robustness defines the regulatory relationship between them. More broadly, learning to predict stochasticity and stability at the physiological and genetic levels from the details of specific error correction strategies of anatomical homeostasis, to flesh out our understanding of the engineering principles that emerge from the evolutionary process [333; 334].
- Cross-level analysis, which extends upwards the ability to decompose ourselves into modules and sub-agents. Tools of information theory [335; 336] or other approaches need to be developed to detect when we are a module in some larger system, to gauge the cognitive capacity of that system, and gain some estimate of what goals are being used by that system to deform the option space available to us locally.
- Exploring radically expanding the kind of system that this framework can handle in terms of embodiment and time-scale. For example, it may be possible to frame evolutionary *lineages* as agents in fitness space [241] and determine how much collective intelligence could be deployed by an evolutionary search process [297; 298; 337]. A promising approach might be to view the recapitulation of phylogenetic stages during development as a kind of memory of the past by the lineage super-agent, as recent work in planaria already showed that other species' morphologies are accessible as other attractors reachable by the bioelectric network of wild-type flatworm tissues [182; 183].
- More broadly, learning to enlarge our cognitive boundary, and gaining an understanding of what an "increased capacity" human (or non-human) would be like, in contrast to the "diminished capacity" with which we are well familiar from legal proceedings. We need to understand how to expand a system's cognitive cone laterally, to care more abundantly [338], as well as upwards, to gain knowledge and therefore responsibility about the larger-scale goals to which our local actions contribute. Overall, a continuum view of agency suggests that important revisions to the concept of responsibility can be developed with respect to collective behavior and "group karma".

*Specific implications: an empirical research program within reach*

While important conceptual issues are advanced over time, the TAME framework already suggests numerous practical research directions immediately within reach (some of which are already pursued in our group):

- Develop models of morphogenetic plasticity and robustness as meta-cognitive error correction mechanisms. For example when the bioelectric layer is shut down in planaria, they stochastically regenerate heads from other species' region of morphospace [182; 183]; when early steps of left-right patterning are disrupted in development, they are often repaired by subsequent steps [339]. One hypothesis is that these are a primitive version of the multiple systems that function in complex brains to correct and oversee each other's actions – the precursor to metacognition. The divergence of current anatomy from bioelectrically-mediated patterns encoding future states to which the organism will repair [184], are perhaps precursors to the advanced ability to consider counterfactual memories and model the future, as well as perceptual bistability [340]. These can now be investigated by bringing the many tools of computational and cognitive neuroscience to bear on non-neural tissues [99].
- Attempt training (to fully understand the cognitive sophistication) and quantify problem-solving capacity in various spaces of a range of diverse models, including cells, tissues, gene-regulatory networks, and organs. When rewarding for morphogenetic outcomes for example, this becomes a direction for regenerative medicine that could bypass the complexity limitations of bottom-up micromanagement of the cellular hardware via molecular medicine. Unlocking the benefits of genomics requires to also understand the software of life and the scaling and control of goal-directed activity at all scales in organisms. By exploiting the endogenous intelligence of cellular swarm agents, we can use techniques further along the spectrum of persuadability than merely rewiring the hardware, and can avoid the inverse problem that limits current approaches to control growth and form. This will give rise to therapies for birth defects, traumatic injury, degenerative disease, aging, and cancer, in which we communicate with cellular collectives and motivate them toward desired anatomical repair.
- As experiments suggest trainability in simple systems like GRNs [106; 341], we should be attempting to train all kinds of networks and systems too complex to micromanage directly – from protein pathways to power grids and beyond. Several mathematical tools including variational approaches, Bayesian inference, and new information theory [119; 326; 327; 342] are now available and ready to be tested.
- Attempt to trigger multicellularity in unicellular species by engineering them to express specific gap junction proteins and ion channels, reconstructing via gain-of-function approaches the events that potentiate multicellularity and the appearance of higher-level selves. Our framework makes numerous predictions about the spread of stress and injury waves within and between organisms, which are testable by modulating these dynamics for quantifiable outcomes in regenerative augmentation.
- Move the bioelectric cancer normalization approach [229; 233] from frogs into mammals for a biomedical roadmap to address cancer based around rescaling the computational boundary of cell networks, not toxic chemotherapy.

- Investigate the cognitive capacities of novel synthetic organisms and smart materials [343; 344; 345; 346] to understand the *origin* of the attractors of behavioral, physiological, and anatomical spaces that are not honed by eons of specific natural selection, linking it to work on "generic" patterns in physical processes [347; 348; 349; 350; 351]. Test those models by predicting the setpoints of synthetic agents that do not have evolutionary histories of selection to serve as the source of their goals and preferences.
- Develop automated tools for estimation of agency – to help human scientists evaluate and relate to systems in the environment. Machine learning or other approaches need to be developed to process behavioral and architecture data to try to predict where one should start on the spectrum of persuadability as an initial guess for a novel system. This would at the same time be a model of animacy detection by cognitive systems [352; 353].
- Exploit the lessons learned about pre-neural and non-neural problem-solving capacities of life in machine learning and AI: create novel non-neuromorphic architectures that could provide approaches to general artificial intelligence that a narrow focus on simulating brain architectures has not produced.
- Robotics and AI can immediately begin exploring multi-scale architectures (of which swarm robotics is just a first step [354]). The consequences will be the first examples of robot cancer (de-coupling malfunctions in which subunits' own local goals supersede the global system goal), but also a great enhancement of our ability to bridge the sim-to-real gap, as the competence of subunits helps accommodate deficiencies of the high-level control signals seen in the real world.

*Beyond basic science: up-to-date ethics*

The TAME framework also has implications for ethics in several ways. The current emphasis for bioethics is on whether bioengineered constructs (e.g., neural cell organoids) are sufficiently like a human brain or not [355], as a criterion for acceptability. TAME suggests that this is insufficient, because many different architectures for cognition are possible (and will be realized) – similarity to human brains is too parochial and limiting a marker for entities deserving of protection and other moral considerations. We must develop a new ethics that recognize the diversity of possible minds and bodies, as what something looks like and how it originated [63; 356] will no longer be a good guide when we are confronted with a myriad of creatures that cannot be comfortably placed within the familiar Earth's phylogenetic tree.

Bioengineering of novel Selves raises our moral responsibility. For eons, humans have been creating and releasing into the world advanced intelligences - via pregnancy and birth of other humans. This, in Dennett's phrase, has been achieved until now via high levels of "competency without comprehension" [357]; however, we are now moving into a phase in which we create beings with comprehension – with rational control over their structure and cognitive capacities, which brings additional responsibility. A new ethical framework will have to be formed without reliance on folk notions such as "machine", "robot", "evolved", "designed", etc. because these categories are now seen to not be crisp natural kinds. Instead, wider approaches (such as Buddhist concern for all sentient beings) may be needed to act ethically with respect to agents that have preferences, goals, concerns, and cognitive capacity in very unfamiliar guises. TAME seeks to break through the biases around contingent properties that drive our estimates of who or what deserves proper treatment, to develop a rational, empirically-based mechanism for recognizing Selves around us.

Another aspect of bioethics is the discussion of limits on technology. Much of it is often driven by a mindset of making sure we don't run afoul of the risks of negative uses of specific technologies (e.g., genetically-modified organisms in ecosystems). This is of course important, with respect to the new bioengineering capabilities. However, such discussions often are one-sided, framed as if the status quo was excellent, and our goal is not to make things worse. This is a fundamental error which neglects the opportunity cost of failing to fully exploit the technologies which could drive advances in the control of biology. The status quo is not perfect – society faces numerous problems including disparities of quality of life across the globe, incredible suffering from unsolved medical needs, climate change, etc. It must be kept in mind that along with the need to limit negative consequences of scientific research, there is a *moral imperative* to advance aspects of this research program that for example will enable the cracking of the morphogenetic code to revolutionize regenerative medicine far beyond what genomic editing and stem cell biology can do alone [358].

The focus on risk arises from a feeling that we should not "mess with nature", as if the existing structures (from anatomical order to ecosystems) are ideal, and our fumbling attempts will disrupt their delicate balance. While being very careful with these powerful advances, it must also be kept in mind that this balance (i.e., the homeostatic goals of systems from cells to species in the food web) was not achieved by optimizing happiness or any other quality commensurate with modern values: it is the result of dynamical systems properties shaped by the frozen accidents of the *random* meanderings of the evolutionary process and the harsh process of selection for survival capacity. Surely we have the opportunity to do better than chance – to use rational design to improve on evolution's harnessing of randomness.

Importantly, current technologies are forcing us to confront an existential risk. Swarm robotics, Internet of Things, AI, and similar engineering efforts are going to be creating numerous complex, goal-driven systems made up of competent parts. We currently have no mature science of where the goals of such novel Selves come from. TAME reminds us that it is essential to understand how goals of composite entities arise and how they can be predicted and controlled. To avoid the Skynet scenario [359], it is imperative to study the scaling of cognition in diverse substrates, so that we can ensure that the goals of powerful, distributed novel beings align with ours.

Given the ability of human subunits to merge into even larger (social) structures, how do we construct higher-order Selves that promote flourishing for all? The multicellularity-cancer dynamic (Figure 9) suggests that tight functional connections that blur cognitive boundaries among subunits is a way to increase cooperation and cognitive capacity. However, simply maximizing loss of identity into massive collectivism is a well-known failure at the social level, always resulting in the same dynamic: the goals of the whole diverge sharply from those of the parts, which become as disposable to the larger social Self as shed skin cells are to us. Thus, the goal of this research program beyond biology is the search for optimal binding policies between subunits, which optimize the tradeoffs needed to maximize individual goals and well-being (preserving freedom) while reaping the benefits of a scaled-up Self at the level of groups and entire societies. While the specific binding mechanisms used by evolution are no guarantee to be the policies we want at the social level, the study of these are critical for jump-starting a rigorous program of research into possible ways of scaling that could have social relevance. These issues have been previously addressed in the context of evolutionary dynamics and game theory [235; 360; 361], but can be significantly expanded using the TAME framework.

In the end, important ethical questions around novel agents made of combinations of hardware, software, evolved, and designed components always come back to the nature of the Self. The coherence of a mind, and its ability to pursue goal-directed activity is central to our notions of moral responsibility in the legal sense: diminished capacity, and soon, enhanced capacity, to make *choices* is a pillar for social structures. Mechanist views of cause and effect in the neuroscience of behavior have been said to erode these classical notions. Rather than reduce Selves (to 0, in some eliminativist approaches), TAME finds novel Selves all around us. We see more agency, not less, when evolution and cell biology are taken seriously [362]. The cognitive Self is not an illusion; what is an illusion is that there is only one, permanent, privileged Self and that it has to arise through the random hill-climbing process of evolution. Our goal, at the biomedical, personal, and social levels is not to destroy the Self but to recognize it in all of its guises, understand its transitions, and enlarge its cognitive capacity toward the well-being of ever more other Selves.

**Acknowledgements**

I thank Christopher Fields, Richard Watson, Eva Jablonka, Joshua Bongard, Pranab Das, Daniel Dennett, Aniruddh Patel, Dora Biro, Rafael Yuste, Giovanni Pezzulo, Thomas Doctor, Bill Duane, Olaf Witkowski, Elizaveta Solomonova, Avery Caulfield, Matthew Simms, Santosh Manicka, Aysja Johnson, Patrick McMillen, Anna Ciaunica, Andrew Reynolds, and numerous others from the Levin Lab and the Diverse Intelligences community for helpful conversations and discussions, as well as comments on versions of this manuscript. I gratefully acknowledge support by the Templeton World Charity Foundation (TWCF0606), the John Templeton Foundation (62212), and The Elisabeth Giauque Trust. This paper is dedicated to my mother, Luba Levin, who while not having been a scientist, always modeled a deep understanding of, and care for, the multi-scale agency abundant in the world.

**Box 1: stress as the glue of agency**

Tell me what you are stressed about and I will know a lot about your cognitive sophistication. Local glucose concentration? Limb too short? Rival is encroaching on your territory? Your limited lifespan? Global disparities in quality of life on Earth? The scope of states that an agent can possibly be stressed by, in effect, defines their degree of cognitive capacity. Stress is a systemic response to a difference between current state and a desired setpoint; it is an essential component to scaling of Selves because it enables different modules (which sense and act on things at different scales and in distributed locations) to be bound together in one global homeostatic loop (toward a larger purpose). Systemic stress occurs when one sub-agent is not satisfied about its local conditions, and propagates its unhappiness outward as hard-to-ignore signals. In this process, stress pathways serve the same function as hidden layers in a network, enabling the system to be more adaptive by connecting diverse modular inputs and outputs to the same basic stress minimization loop. Such networks scale stress, but stress is also what helps the network scale up its agency – a bidirectional positive feedback loop.

The key is that this stress signal is unpleasant to the other sub-agents, closely mimicking their own stress machinery (genetic conservation: my internal stress molecule is the same as your stress molecule, which contributes to the same "wiping of ownership" that is implemented by gap junctional connections). By propagating unhappiness in this way (in effect, turning up the global system "energy" which facilitates tendency for moving in various spaces), this process recruits distant sub-agents to act, to reduce their own perception of stress. For example, if an organ primordium is in the wrong location and needs to move, the surrounding cells are more willing to get out of the way if by doing so they reduce the amount of stress signal they receive. It may be a process akin to run-and-tumble for bacteria, with stress as the indicator of when to move and when to stop moving, in physiological, transcriptional, or morphogenetic space. Another example is compensatory hypertrophy, in which damage in one organ induces other cells to take up its workload, growing or taking on new functions if need be [363; 364]. In this way, stress causes other agents to work toward the same goal, serving as an influence that binds subunits across space into a coherent higher Self and resists the "struggle of the parts" [365]. Interestingly, stress spreads not only horizontally in space (across cell fields) but also vertically, in time: stress response is one of the things most easily transferred by transgenerational inheritance [366].

**Figure Legends**

Figure 1: Diverse, Multiscale Intelligence

(**A**) Biology is organized in a multi-scale, nested architecture of molecular pathways. (**B**) These are not merely structural, but also computational: each level of this holarchy contains subsystems which exhibit some degree of problem-solving (i.e., intelligent) activity, on a continuum such as the one proposed by Rosenblueth et al. [4]. (**C**) At each layer of a given biosystem, novel components can be introduced of either biological or engineered origin, resulting in chimeric forms that have novel bodies and novel cognitive systems distinct from the typical model species on the Earth's phylogenetic lineage. Images in panels A,C by Jeremy Guay of Peregrine Creative. Image in panel B was created after [4].

Figure 2: the Axis of Persuadability

A proposed way to visualize a continuum of agency, which frames the problem in a way that is testable and drives empirical progress, is via an "axis of persuadability": to what level of control (ranging from brute force micromanagement to persuasion by rational argument) is any given system amenable, given the sophistication of its cognitive apparatus? Here are shown only a few representative waypoints. On the far left are the simplest physical systems, e.g. mechanical clocks (**A**). These cannot be persuaded, argued with, or even rewarded/punished – only physical hardware-level "rewiring" is possible if one wants to change their behavior. On the far right (**D**) are human beings (and perhaps others to be discovered) whose behavior can be radically changed by a communication that encodes a rational argument that changes the motivation, planning, values, and commitment of the agent receiving this. Between these extremes lies a rich panoply of intermediate agents, such as simple homeostatic circuits (**B**) which have setpoints encoding goal states, and more complex systems such as animals which can be controlled by signals, stimuli, training, etc. (**C**). They can have some degree of plasticity, memory (change of future behavior caused by past events), various types of simple or complex learning, anticipation/prediction, etc. Modern "machines" are increasingly occupying right-ward positions on this continuum [63]. Some may have preferences, which avails the experimenter of the technique of rewards and punishments – a more sophisticated control method than rewiring, but not as sophisticated as persuasion (the latter requires the system to be a logical agent, able to comprehend and be moved by arguments, not merely triggered by signals). An important great transition is becoming sophisticated enough to be susceptible to a "thought that breaks the thinker" (e.g., existential or skeptical arguments that can make one depressed or even suicidal, Gödel paradoxes, etc.) – massive changes can be made in those systems by a very low-energy signal because it is treated as information in the context of a complex host computational machinery. These agents exhibit a degree of multi-scale plasticity that enables informational input to make strong changes in the structure of the cognitive system itself. The positive flip side of this vulnerability is that it avails those kinds of minds with a long term version of free will: the ability through practice and repeated effort to change their own thinking patterns, responses to stimuli, and functional cognition. This continuum is not meant to be a linear *scala naturae* that aligns with any kind of "direction" of evolutionary progress – evolution is free to move in any direction in this option space of cognitive capacity; instead, this scheme provides a way to formalize (for a pragmatic, engineering approach) the major transitions in cognitive capacity that can be exploited for increased insight and control. The goal of the scientist is to find the optimal position for a given system. Too far to the right, and one ends up attributing hopes and dreams to thermostats or simple AIs in a way that does not advance prediction and control. Too far to the left, and one loses the benefits of top-down control in favor of intractable micromanagement. Note also that this forms a continuum with respect to how much knowledge one has to have about the system's details in order to manipulate its function: for systems in class A, one has to know a lot about their workings to modify them. For class B, one has to know how to read-write the setpoint information, but does not need to know anything about how the system will implement those goals. For class C, one doesn't have to know how the system modifies its goal encodings in light of experience, because the system does all of this on its own – one only has to provide rewards and punishments. Images by Jeremy Guay of Peregrine Creative.

Figure 3: Cognitive Selves can change in real-time
(**A**) Caterpillars metamorphose into butterflies, going through a process in which their body, brain, and cognitive systems are drastically remodeled during the lifetime of a single agent. Importantly, memories remain and persist through this process [367]. (**B**) Planaria cut into pieces regenerate, with each piece re-growing and remodeling precisely what is needed to form an entire animal. (**C**) Planarians derived from tail fragments of trained worms still retain original information, illustrating the ability of memories to move across tissues and be reimprinted on newly-developing brains [19; 20; 368]. Images by Jeremy Guay of Peregrine Creative.

Figure 4: Unconventional goal-directed agents and the scaling of the cognitive Self
(**A**) The "hydrogen atom" of agency is homeostasis, for example the ability of a cell to execute the Test-Operate-Exit [161] loop: a cycle of comparison with setpoint and adjustment via effectors, which allows it to remain in a particular region of state space. (**B**) This same capacity is scaled up by cellular networks into anatomical homeostasis: morphogenesis is not simply a feedforward emergent process but rather the ability of living systems to adjust and remodel to specific target morphologies. This requires feedback loops at the transcriptional and biophysical levels, which rely on stored information (e.g., bioelectrical pattern memories) against which to minimize error. (**C**) This is what underlies complex regeneration such as salamander limbs, which can be cut at any position and result in just the right amount and type of regenerative growth that stops when a correct limb is achieved. Such homeostatic systems are examples of simple goal-directed agents. (**D**) A focus on the size or scale of goals any given system can pursue allows plotting very diverse intelligences on the same graph, regardless of their origin or composition [110]. The scale of their goal-directed activity is estimated (collapsed onto one axis of space and one of time, as in Relativity diagrams). Importantly, this way of visualizing the sophistication of agency is a schematic of goal space – it is not meant to represent the spatial extent of sensing or effector range, but rather the scale of events about which they care and the boundary of states that they can possibly represent or work to change. This defines a kind of cognitive light cone (a boundary to any agent's area of concern); the largest area represents the "now", with fading efficacy both backward (accessing past events with decreasing reliability) and forward (limited prediction accuracy for future events). Agents are compound entities, comprised of (and comprising) other sub- or super-agents each of which has their own cognitive boundary of various sizes. Images by Jeremy Guay of Peregrine Creative.

Figure 5: Cognitive agents solve problems in diverse spaces
	Intelligence is fundamentally about problem-solving, but this takes place not only in familiar 3D space as "behavior" (control of muscle effectors for movement) (**A**), but also in other spaces in which cognitive systems try to navigate, in order to reach better regions. This includes the transcriptional space of gene expression (**B**) here schematized for two genes, anatomical morphospace (**C**) here schematized for two traits, and physiological space (**D**) here schematized for two parameters. An example (**E**) of problem-solving is planaria, which placed in barium (causing their heads to explode due to general blockade of potassium channels) regenerate new heads that are barium-insensitive [141]. They solve this entirely novel (not primed by evolutionary experience with barium) stressor by a very efficient traversal in transcriptional space to rapidly up/down regulate a very small number of genes that allows them to conduct their physiology despite the essential $K^+$ flux blockade. (**F**) The degree of intelligence of a system can be estimated by how effectively they navigate to optimal regions without being caught in a local maximum, illustrated as a dog which could achieve its goal on the other side of the fence, but this would require going around - temporarily getting further from its goal (a measurable degree of patience or foresight of any system in navigating its space, which can be visualized as a sort of energy barrier in the space, **F'**).  Images by Jeremy Guay of Peregrine Creative

Figure 6: Morphogenesis as an example of collective intelligence and plasticity
	The results of complex morphogenesis are the behavior in morphospace of a collective intelligence of cells. It is essential to understand this collective intelligence because by themselves, progress in molecular genetics is insufficient. For example, despite genomic information and much pathway data on the behavior of stem cells in planarian regeneration, there are no models predicting what happens when cells from a flat-headed species are injected into a round-headed species (**A**): what kind of head will they make, and will regeneration/remodeling ever stop, since the target morphology can never match what either set of cells expects?  Development has the ability to overcome unpredictable perturbations to reach its goals in morphospace: tadpoles made with scrambled positions of craniofacial organs can make normal frogs (B) because the tissues will move from their abnormal starting positions in novel ways until a correct frog face is achieved [174]. This illustrates that the genetics specifies not an invariant set of movements but an error minimization (homeostatic) loop with reference to a stored anatomical setpoint (target morphology). The paths through morphospace are not unique, illustrated by the fact that when frog legs are induced to regenerate (**C**), the intermediate stages are not like the developmental path of limb development (forming a paddle and using programmed cell death to separate the digits) but rather like a plant (**C'**), in which a central core gives rise to digits growing as offshoots (green arrowheads) which nevertheless ends up being a very normal-looking frog leg [176]. The plasticity extends across levels: when newt cells are made very large by induced polyploidy, they not only adjust the number of cells that work together to build kidney tubules with correct lumen diameter, but can call up a completely different molecular mechanism (cytoskeletal bending instead of cell:cell communication) to make a tubule consisting in cross-section of just 1 cell wrapped around itself; this illustrates intelligence of the collective, as it creatively deploys diverse lower-level modules to solve novel problems. The plasticity is not only structural but functional: when tadpoles are created to (**E**) have eyes on their tails (instead of in their heads), the animals can see very well [38], as revealed by their performance in visual learning paradigms (**F**). Such eyes are also competent modules: they first form correctly despite their aberrant neighbors (muscle, instead of brain), then put out optic nerves which they connect to the nearby spinal cord, and later they ignore the programmed cell death of the tail, riding it backwards to end up on the posterior of the frog (**G**). All of this reveals the remarkable multi-scale competency of the system which can adapt to novel configurations on the fly, not requiring evolutionary timescales for adaptive functionality (and providing important buffering for mutations that make changes whose disruptive consequences are hidden from selection by the ability of modules to get their job done despite changes in their environment).

Figure 7: Bioelectrical pattern memories

Planarian fragments reliably regenerate whatever is missing, and stop when a correct worm is complete. Normal planaria (A) have 1 head and 1 tail (A-1), expression of anterior genes in the head (A-2), and a standing pattern of resting potential that depolarized on the end that should make a head (A-3, revealed by voltage-reporting fluorescent dye, depolarized region marked with orange arrowhead). When a middle portion is amputated, it regenerates to a correct 1-headed worm (A-4). It is possible to edit the information structure which encodes the target morphology (shape to which the fragments will regenerate). In worms that are anatomically normal (A'-1), with normal gene expression (A'-2), their bioelectric pattern can be altered in place (A'-3, orange arrowheads mark the two depolarized ends) using ion channel-targeting drugs or RNAi. The result, after injury, will be a fully viable 2-headed worm (A'4). Importantly, the pattern in A'-3 is not a voltage map of the final 2-headed worm: it's a map of a 1-headed animal before cutting, which already has the induced false memory indicating that a correct worm should have 2 heads. This information is latent, only guiding the cellular collective's anatomical homeostasis activity after injury. Thus it is also a basal example of counterfactual representation, referring to what should happen *if* an injury occurs, not what is happening now. Such changes to the bioelectric target morphology are true memories because they are re-writable but also long-term stable: if cut again, in water with no more channel-perturbing reagents, multiple rounds of regeneration of a genetically wild-type worm continue to give rise to 2-headed forms (B), which can be re-set back to normal by a different bioelectric perturbation [41]. The control of morphology by bioelectric patterns is mediated as in the brain (C) by cells which have ion channels that set resting potential across the membrane ($V_{mem}$) and propagate those states in computational networks to their neighbors, via electrical synapses known as gap junctions. All cells, not just neurons (D) do this, and bioelectric signaling is an ancient information processing modality that pre-dates neurons and brains [181; 204]. The ability of voltage states to functionally specify modular anatomy is seen when an ion channel is used to set membrane voltage in endodermal cells fated to be gut, to an eye-like bioelectric prepattern (E), which then create an eye on the gut (red arrowhead) [260]. This phenomenon has 2 levels of instruction (F): in addition to our use of voltage to instruct shape at the organ level (not micromanaging individual eye components), the ion channel mRNA-injected cells (cyan β-galactose marker) further instruct their neighbors (brown cells) to participate in forming this ectopic lens. Images in panels C,D are courtesy of Peregrine Creative. Images in panels A and A' are taken with permission from [184]. Embryo image in panel E is used with permission from [369].

Figure 8: Scaling of computation in cells

Individual cells (A) have a degree of computational capacity consisting of the ability to sense local microenvironment, and some memory and ability to anticipate into the future. When assembling into networks (A'), tissues acquire the ability to sense and act at greater spatial distance, as well as gain larger capacity for memory and prediction via greater computational capacity. As neural networks use hidden layers to abstract patterns in data and recognize meso-scale features (B), tissue networks gain the capacity to represent information larger than the molecular and cell level: each cell's activity (differentiation, migration, etc.) can be the result of other layers of cells processing information about current and past states, enabling decision-making with respect to tissue, organ, or whole organism-scale anatomy. (C) Much as some neural networks store individual memories as attractors in their state space, bioelectric circuits' attractors function as pattern memories, triggering cells to execute behaviors that implement anatomical outcomes like number and location of heads in planaria. Images courtesy of Jeremy Guay of Peregrine Creative.

Figure 9: Gap junctions and the cellular collective

Communication via diffusible and biomechanical signals can be sensed by receptors at the membrane as messages coming from the outside of a cell (A). In contrast, cells coupled by gap junctions enable signals to pass directly from one cell's internal milieu into another. This forms a partial syncytium which helps erase informational boundaries between cells, as memory molecules (results of pathway dynamics) propagate across such cell groups without metadata on which cell originated them. The versatile gating of GJ synapses allows the formation of multicellular Selves that own memories of physiological past events at the tissue level (not just individual cells') and support larger target patterns, enabling them to cooperate to make complex organs (B). This process can break down: when oncogenes are expressed in tadpoles, voltage dye imaging (C) reveals the abnormal voltage state of cells that are disconnected bioelectrically from their neighbors, reverting to a primitive unicellular state (metastasis) that treats the rest of the body as external environment and grows out of control as tumors (D). This process can be prevented [232; 275] by artificially regulating their bioelectric state (e.g., co-injecting a hyperpolarizing channel with the oncogene, E). In this case the tissue forms normally (F, green arrow), despite the very strong presence of the oncogene (G, red label). This illustrates the instructive capacity of bioelectric networks to dominate single cell and genetic states to control large-scale tissue outcomes.

Figure 10: Gap junctions scale homeostatic goals

Gap junctions are a type of connection architecture that facilitates scaling of goal states (and thus expands the cognitive cone of cellular agents). A single cell's homeostatic cycle has 3 parts: measurements, comparison to a memory setpoint, and acting via effectors to stay in or reach the correct region of state space (A). When coupled into gap junctions (B), each of these 3 components expands in size and complexity: the cell group is able to 1) measure a larger region (reacting to spatially more complex inputs, not just local conditions), 2) store a more complex setpoint pattern, and 3) act (deform, grow, etc.) at a scale that produces large-scale anatomical change. The goals of such networks readily map on to regeneration and regulative development (C): dynamical systems pictures of artificial neural networks as they perform pattern completion based on partial input illustrate an energy landscape with wells corresponding to stable target morphology memories. The process of completing a correct planarian pattern from a simple fragment can be modeled in this way, perhaps with overall stress levels instantiating the free energy that the system is trying to minimize [342]. Such attractors correspond to different possible morphologies, and indeed the normally robust regeneration toward a single pattern (D) can be modified in planaria by temporarily disrupting their gap junctional network, which causes genetically un-modified worms to nevertheless build heads appropriate to other species' attractors in morphospace (E) [182; 183]. Images in A,B courtesy of Jeremy Guay of Peregrine Creative. Images in C,D,E courtesy of Alexis Pietak.

Figure 11: Technology reveals gradualism in Descartes' cut

The apparent fundamental gulf between first person perspective (what is it like to *be* a specific Self) and third person perspective (external scientific study of that cognitive system) can be seen to also be a gradual continuum, when modern technology is used to expand heterophenomenology [370]. On the left side of the continuum is a traditional 3-rd person scenario of an agent studying another by measuring its physical states: cognitive states can be inferred but not directly experienced (correlates of consciousness), giving rise to the problem of other minds, and a firm distinction between "you" and "me". However, sensory substitution and augmentation technology now enables the plugging of various peripherals and sensors directly into the nervous system of subjects, and thus it is possible to connect the output of electrophysiology equipment directly into the subject's brain. For example if the output of a multielectrode array recording neural activity of subject #1 is connected directly to the brainport device [45] of subject #2, allowing #1's mental states to more directly provide input into #2's sensory stream. This can be made even more direct by fusing portions of two brains directly, during embryogenesis, illustrating that the strength of boundaries between "you" and "me" is variable. It's critical to note that these fusion experiments are not just aberrant corner cases because all brains are already fusions of neural modules. Single subject's brains consist of two hemispheres which must communicate to give rise to a coherent, centralized perception of "me" despite being made of communicating parts, and can be dissociated. Indeed, all the parts of the brain are fused together already into an architecture that gives rise to a single subjective Self, and this kind of fusion can be reproduced or expanded upon to whatever degree necessary by biological or technological fusion among subjects. Importantly however, what happens when one fuses cognitive systems with their subject of study is that a new Self appears (a composite cognitive system), showing that the Self can remain invariant while pursuing scientific study of functional cognition and behavior (the left of the spectrum), but essentially must change in order to gain first-hand knowledge of consciousness in other cognitive systems. Images are courtesy of Jeremy Guay of Peregrine Creative.

**Tables**

Table 1: the core tenets of TAME
- Continuum of cognitive capacities – no binary categories, no bright line separating true cognition from "just physics", as is clear from evolutionary process and ability to bioengineer chimeras between any two "natural kinds"
- Frameworks must apply to truly diverse intelligences – beyond the examples from Earth's phylogenetic tree based on brains, we must be able to consider and compare agents across the option space of designed and evolved combinations of living, non-living, and software components at all scales.
- Selves exist across a continuum of persuadability, and it is an empirical question as to where on this axis any given system lies (revealed by the ratio of prediction and control vs. effort and knowledge that needs to be input, for any given way of relating to that system).
- Selves are not fixed, permanent agents – their substrate can remodel radically during their lifetime; the owner of memories and preferences, and the subject of rewards and punishments, is malleable and plastic.
- The core of being a Self is the ability to pursue goals. Selves can be nested and overlapping, cooperating and competing both laterally and across levels. Each higher-level self deforms the option space for the lower level Selves, enabling them to follow energy minimization to achieve outcomes that look inevitable and simple at one scale, while serving intelligent goals at a higher scale.
- Intelligence is the degree of competency of navigating any space (not just 3D space), including morphospace, transcriptional space, physiological space, etc. toward desirable regions, while avoiding being trapped in local minima.

Table 2: isomorphism between cognition and pattern formation

| Cognitive concept | Morphogenetic concept |
|---|---|
| Patterns of activation across neural networks processing information | Differential patterns of $V_{mem}$ across tissue formed by propagation of bioelectric states through gap junction synapses. |
| Local field potential (EEG) | $V_{mem}$ distribution of cell group |
| Intrinsic plasticity | Change of ion channel expression based on $V_{mem}$ levels |
| Synaptic plasticity | Change of cell:cell connectivity via $V_{mem}$'s regulation of gap junctional connectivity |
| Activity-dependent transcriptional changes | Bioelectric signals' regulating gene expression during patterning |
| Neuromodulation, and neurotransmitters controlled by electrical dynamics to regulate genes in neurons | Developmental (pre-nervous) signaling via the same neurotransmitters (e.g. serotonin) moving under control of bioelectrical gradients to regulate second messenger pathways and gene expression. |
| Direct transmission | Cell:cell sharing of voltage via nanotubes or gap junctions |

| Volume transmission | Cell:cell communication via ion levels outside the membrane or voltage-dependent neurotransmitter release |
|---|---|
| Synaptic Vesicles | Exosomes |
| Sensitization | Cells become sensitized stimuli, such as for example to BMP antagonists during development |
| Functional lateralization | Left-right asymmetry of body organs |
| Taste and olfactory perception | Morphogenetic signaling by diffusible biochemical ligands |
| Activity-dependent modification of CNS | Control of anatomy by bioelectric signaling within those same cells |
| Critical plasticity periods | Competency windows for developmental induction events |
| Inborn behaviors (instincts) | Emergent morphogenetic cascades as "default" outcomes of a genetically-specified bioelectric hardware - hardwired patterning programs (mosaic development |
| Voluntary movement | Remodeling, regeneration, metamorphosis |
| Memory | Shorter range: Regeneration of specific body organs. Longer range: Morphological homeostasis over decades as individual cells senesce; altering basic body anatomy in planaria by direct manipulation of bioelectric circuit |
| Counterfactual memories | Ability of 1-headed planarian bodies to store bioelectric patterns indicative of 1-headed or 2-headed forms, which are latent memories that become instructive upon damage to the organism. |
| Perceptual Bistability | Cryptic Planaria, induced by gap-junctional disruption, fragments of which stochastically regenerate as 1-headed or 2-headed forms, shifting between two different bioelectrical representations of a target morphology (pattern memory). |
| Pattern completion ability of neural networks (e.g., attractor nets) | Regeneration of missing parts in partial fragments (e.g., planaria, salamander appendages, etc.) |
| Forgetting | Degradation of target morphology setpoint information leading to cancer and loss of regenerative ability |
| Addiction | Dependency on cellular signals, such as nerve addiction in limb regeneration and cancer addiction to specific molecules. |
| Encoding | Representation of patterning goal states by bioelectric properties of tissue |
| Visual system feature detection | Organ-level monitoring of body configuration and detection of specific boundaries by tissue (such as the $V_{mem}$ boundary that drives brain morphogenesis) |
| Holographic (distributed) storage | Any small piece of a planarian remembers the correct pattern (even if it has been re-written) |

| Behavioral plasticity | Regulative developmental programs and regenerative capacity |
| Self-modeling | Representations of current and future morphogenetic states by bioelectric patterns such as the planarian prepattern or the bioelectric face pattern in vertebrates |
| Goal-seeking | Embryogenesis and regeneration work towards a specific target configuration despite perturbations |
| Adaptivity and Intelligence | Morphological rearrangements carry out novel, not hardwired, movements to reach the same anatomical configuration despite unpredictable initial starting state |
| Age-dependent cognitive decline | Age-dependent loss of regenerative ability |
| Top-down control | Place conditioning for drug effects – top-down control of signaling pathways |

*Legend:* possible mapping of concepts in cognitive neuroscience to examples in pattern formation.

Table 3: An example of the scaling of cognition
- Each Self has a cognitive capacity defined by the spatial, temporal, and complexity metrics on the goals it can possibly pursue.
- Biological Selves scale up by cells' joining into computational networks that can pursue larger-scale (anatomical, not just metabolic) goals.
- Networks increase the spatial reach of sensing and actuation, and increase the computational capacity which allows scaling up of goals and of the states that can induce stress.
- Bodies consist of components which are themselves competent (goal-seeking modules that navigate their own spaces) and can achieve specific outcomes despite perturbations and changing conditions.
- Gap junctions are a unique scaling mechanism which, by linking cells' internal milieus, wipes ownership information on signaling molecules. This partially erases the informational identity of the cellular subunits, driving up cooperation and resulting in novel tissue and organ-level Selves with morphological-scale goals.
- Bioelectric networks underlie the computations of cell collectives at the tissue, organ, and organism scale, propagating stress information, state sensing, and morphogenetic instructive cues over larger areas.
- Selves can dissociate (scale down), as occurs in cancer, by shrinking the computational boundaries of some subunits that de-couple from the network.

Table 4: evolution benefits from multiscale competency of components
- Attractors in the developmental modules' space (their goals) smooth the fitness landscape by reducing phenotypic differences between different genotypes. A kind of Baldwin effect arises from the physiological layer's nano-cognitive capacity that can improve fitness while the genetics catches up at a slower time scale.
- Ability to make up for changes in the microenvironment reduces apparent pleiotropy of the system by enabling consequences of mutation to be examined that would otherwise be masked by maladaptive other consequences of the same mutation. Evolution doesn't have to wait until a solution is found that improves one property while not impairing others.
- Tolerance of mutations due to homeostatic repair enables evolution to explore novel opportunities, not only solve problems.
- Relationship between genotype and phenotype becomes more linear and easier to control, via combination of top-down representation of goal states and bottom-up emergence. This in turn allows much more efficient functional architecture, as inverse problem is minimized.
- Credit assignment during evolution is facilitated, and higher-level Selves can work in an easier (lower-dimensional) spaces such as reward space for subunits, instead of needing to evolve solutions for all of the micro-scale details. This raises the apparent IQ of the evolutionary search process itself, helping it better navigate the fitness landscape space.
- Bioelectric networks provide modularity (triggers of complex subroutines), and coarse-graining (since voltage is a high-order parameter over ion channel gene and protein microstates, and individual ion concentrations).

- Bioelectric networks reduce the limitations of the inverse problem, by inserting a physiological layer between the genome and the anatomy. The relationships of genes to bioelectric patterns, and bioelectric patterns to anatomy, are much easier to reverse and control than the emergent relationship of genome to anatomy directly. This not only helps natural evolution but also helps engineers design and evolve complex proto-cognitive circuits

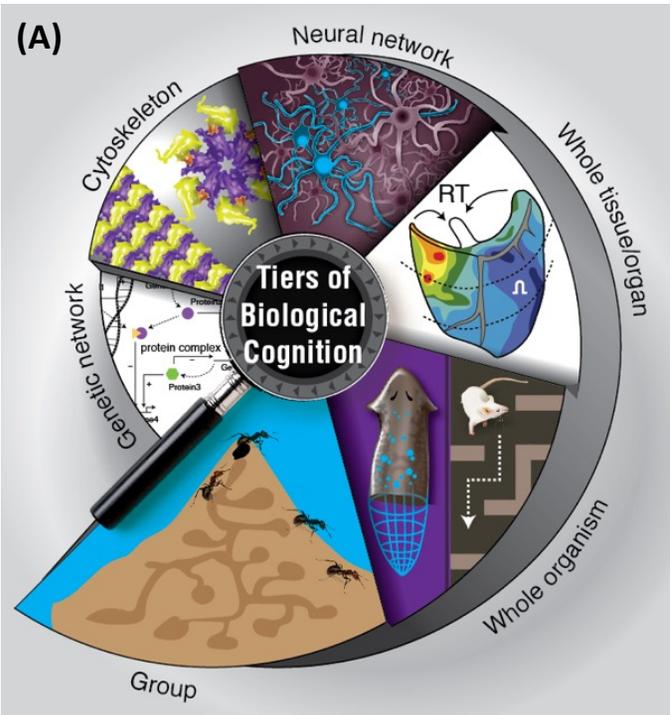
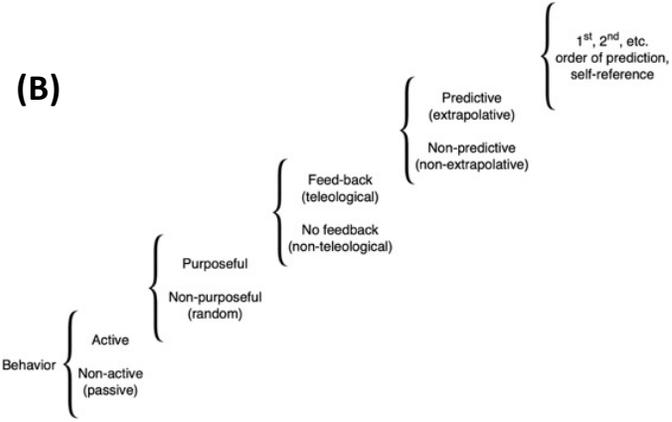
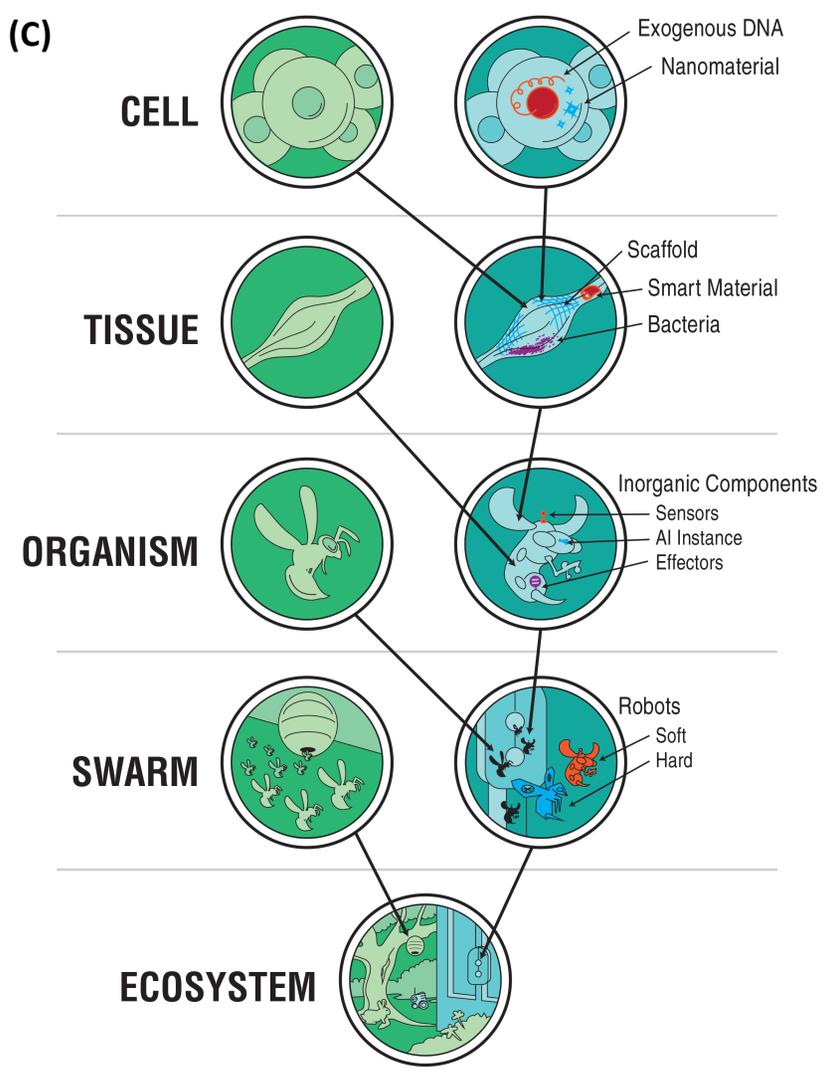

Levin, Figure 1

**(A)** 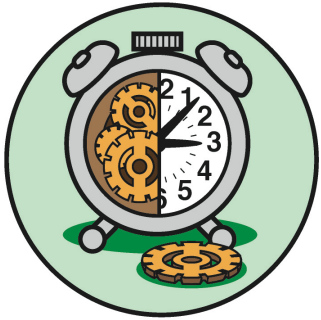
Hardware modification only

**(B)** 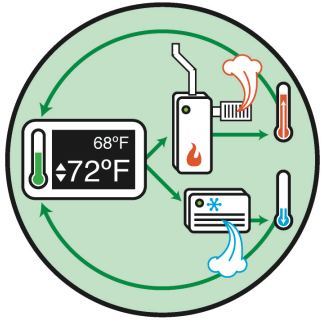
Modify the data encoding setpoint of goal-driven process

**(C)** 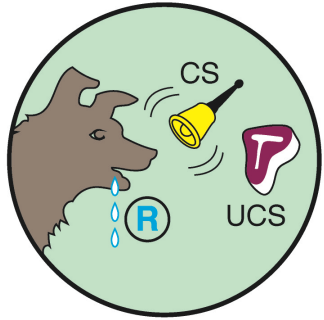
Training by rewards/ punishments

**(D)** 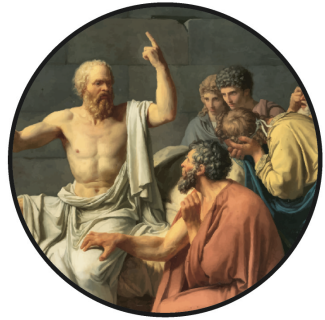
Communicate cogent reasons

Persuadability →

← Effort Needed to Exert Influence

← Mechanism Knowledge Needed to Exert Influence

Levin, Figure 2

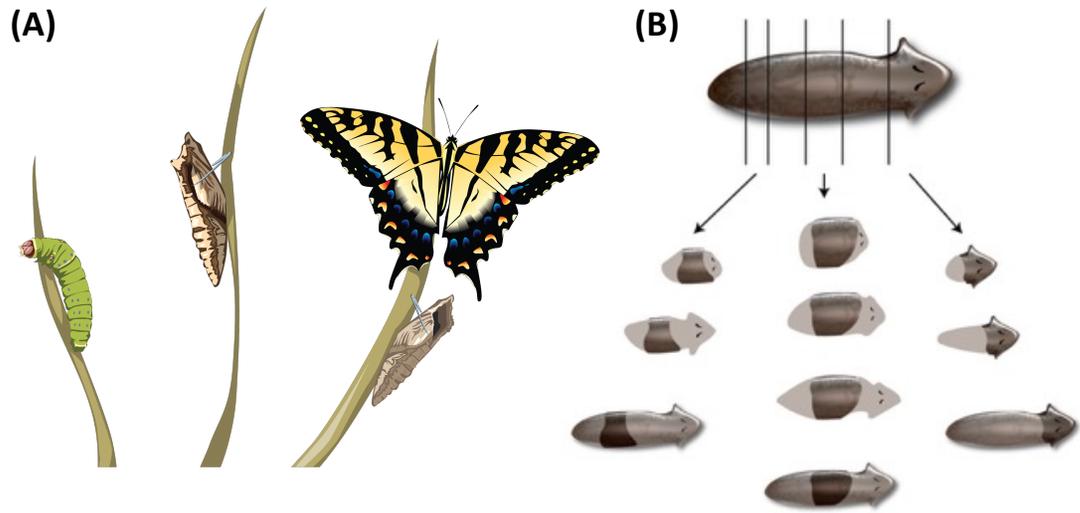
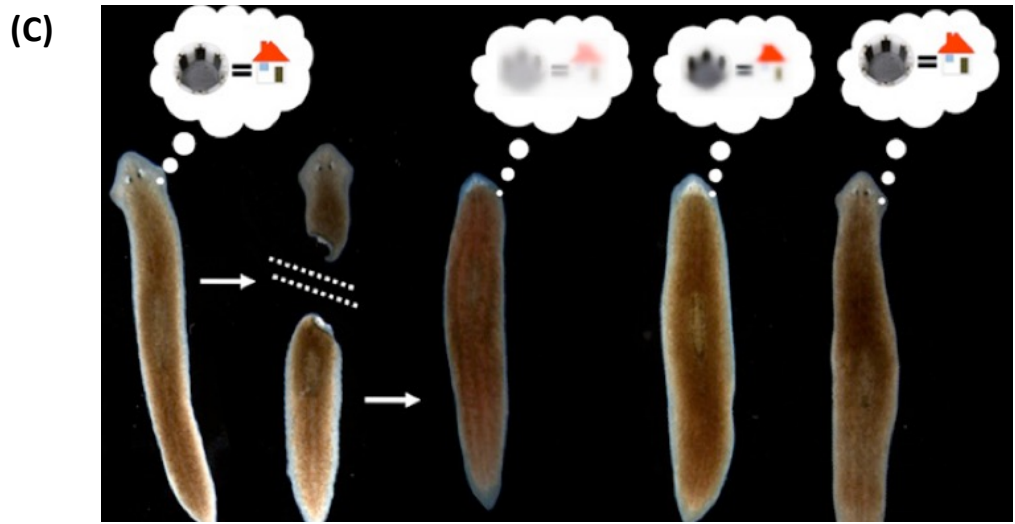

Levin, Figure 3

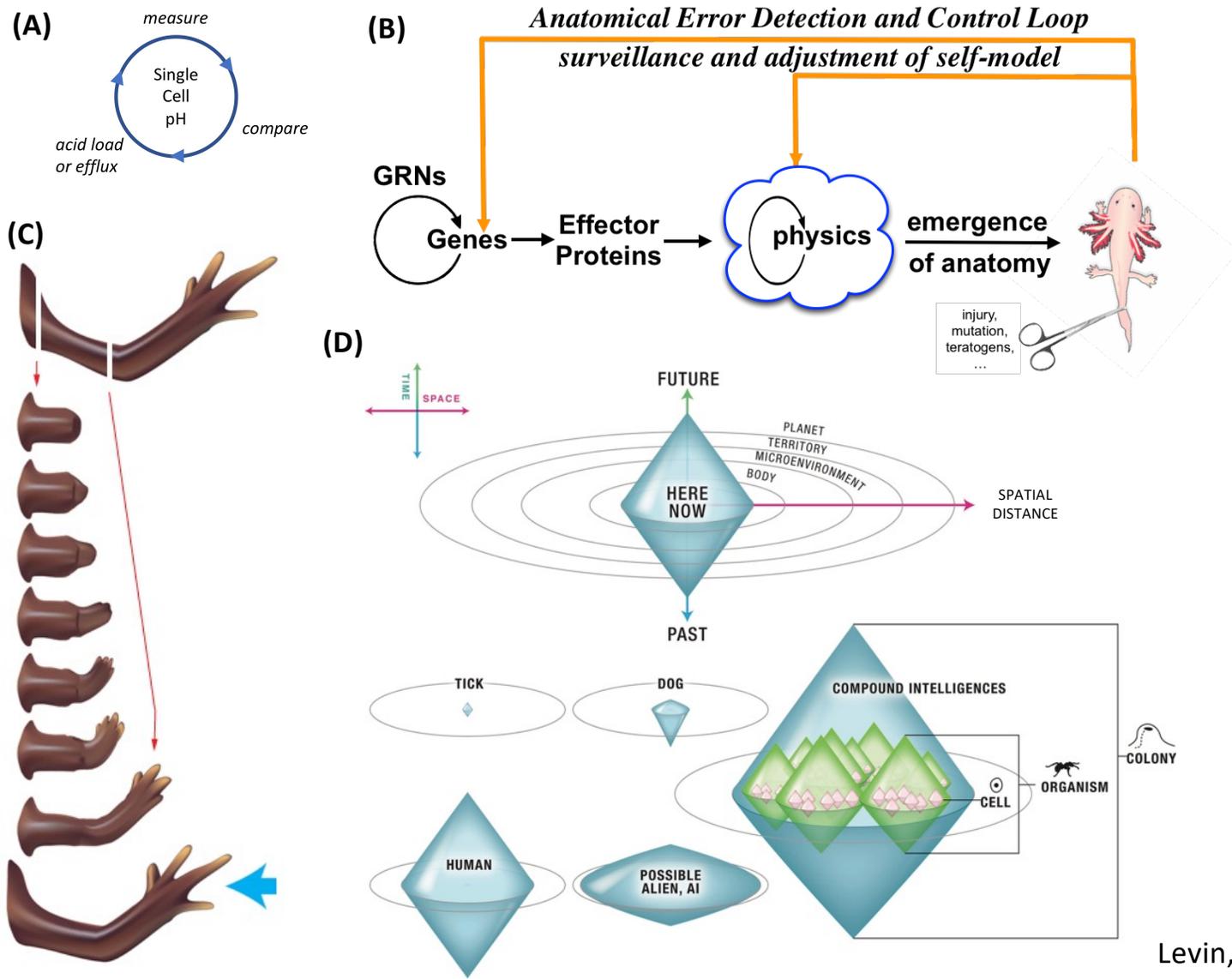

Levin, Figure 4

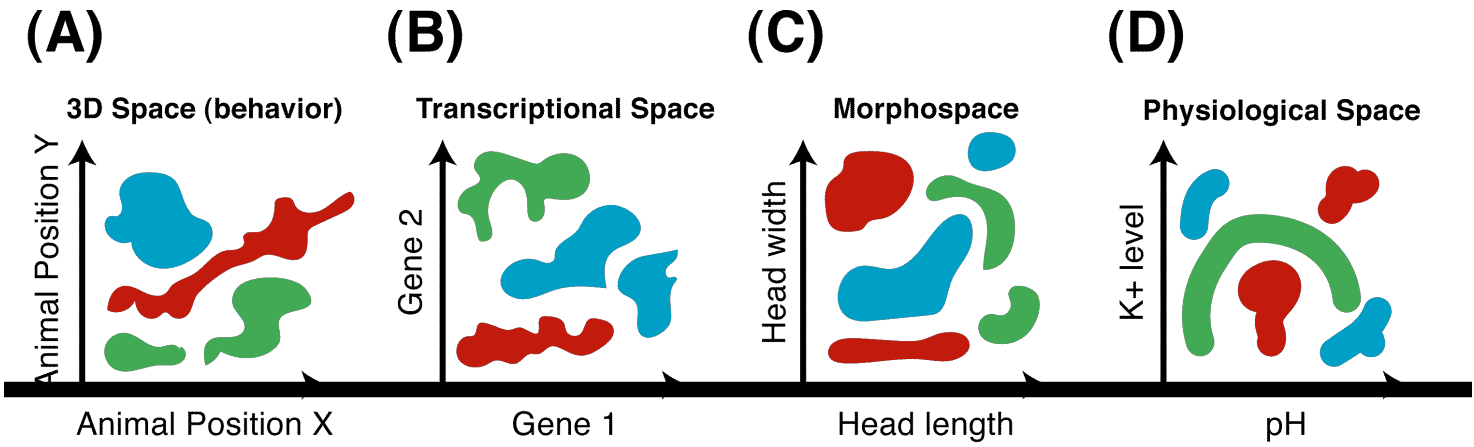
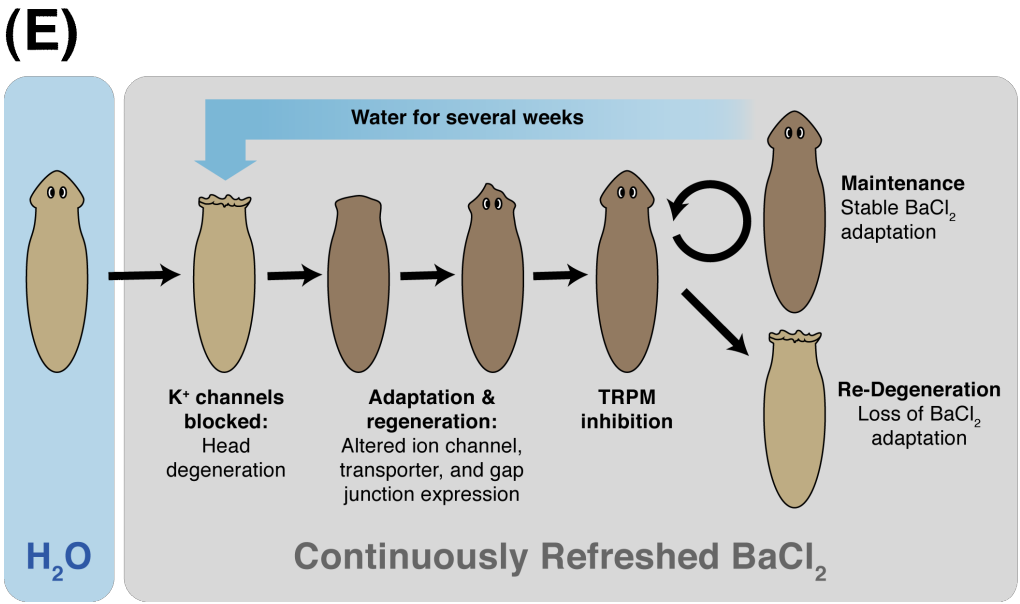
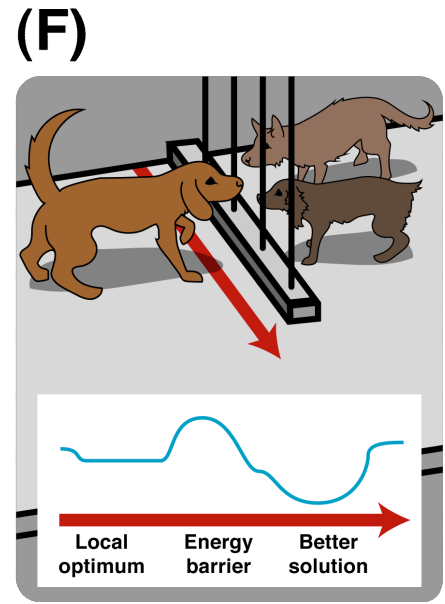

Levin, Figure 5

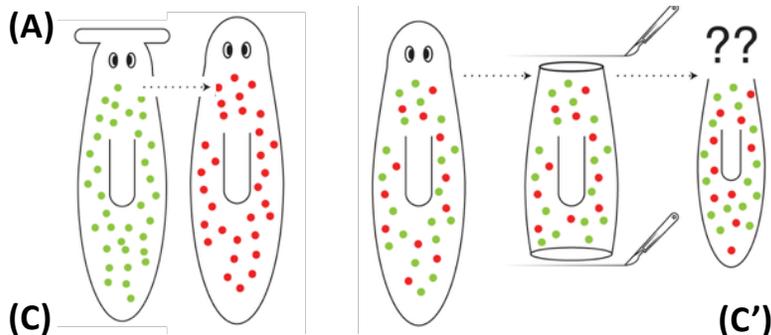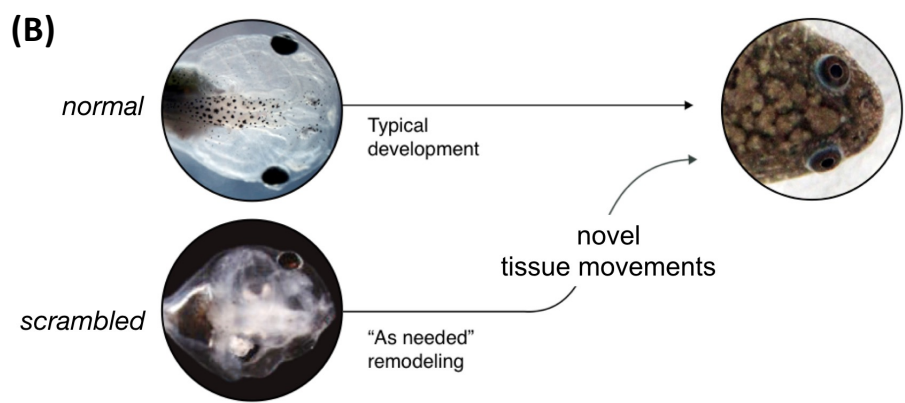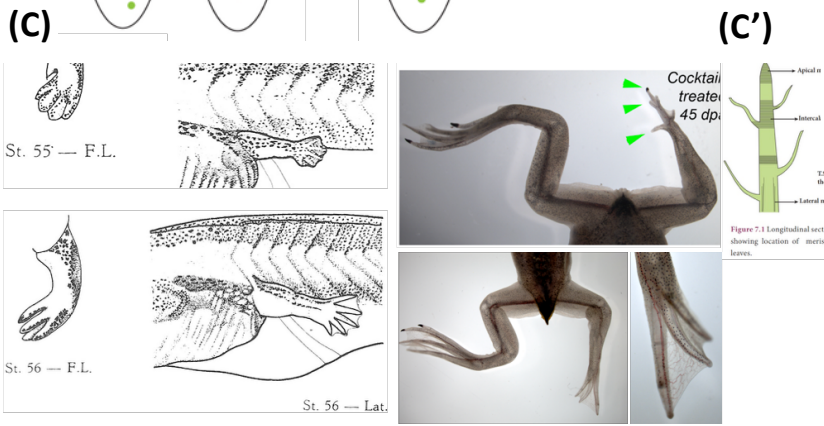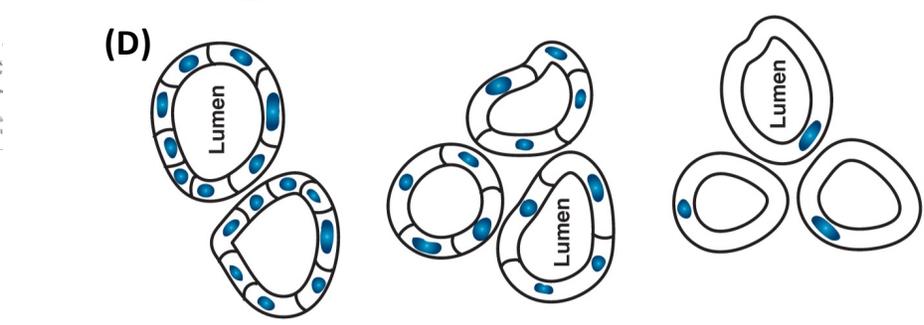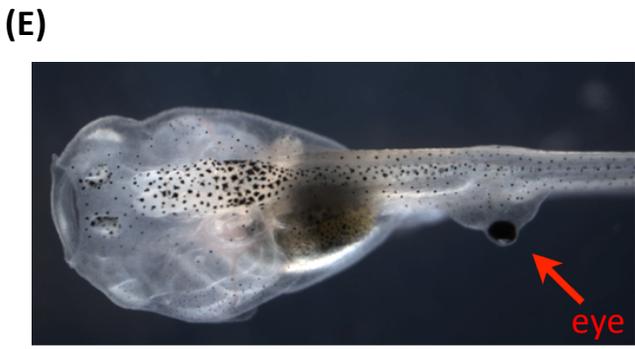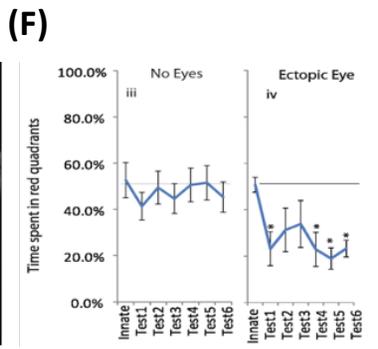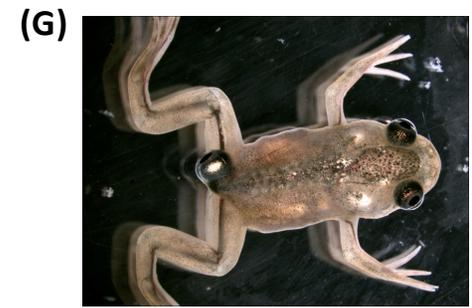

Levin, Figure 6

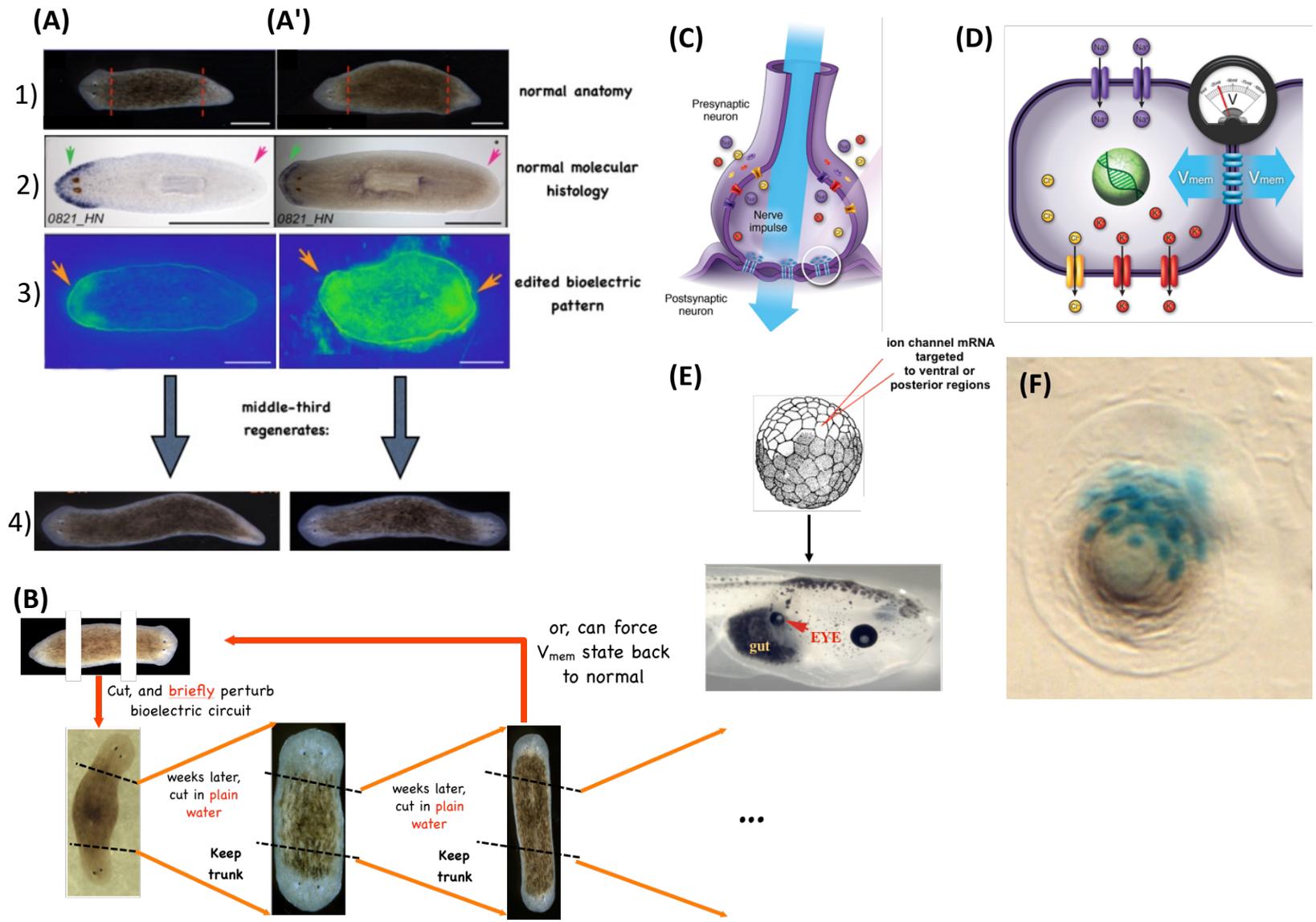

Levin, Figure 7

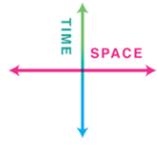
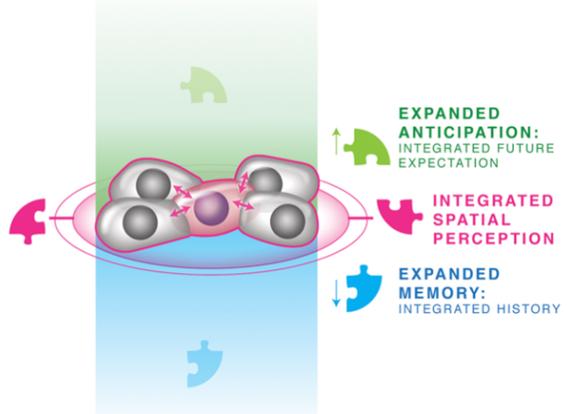
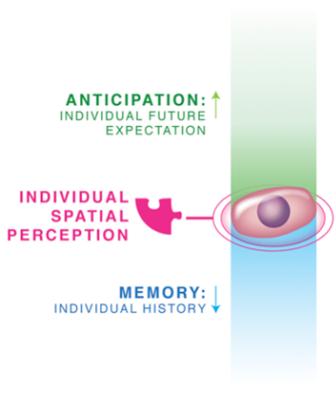
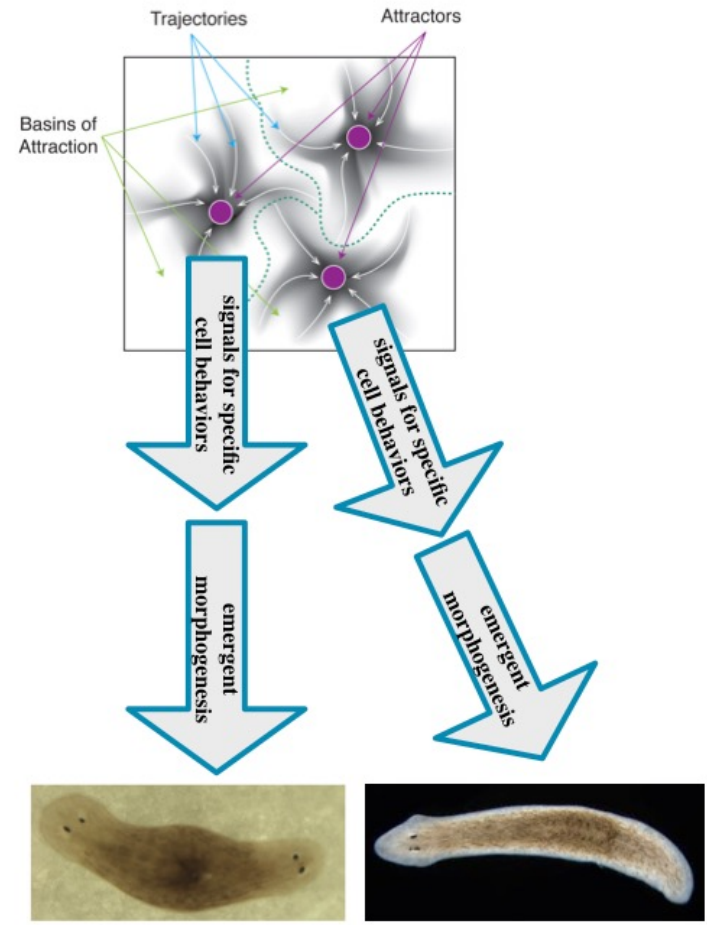
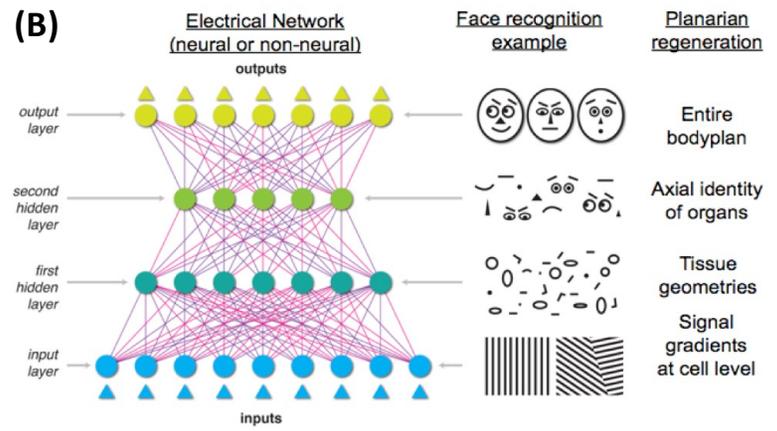

Levin, Figure 8

**(A)** 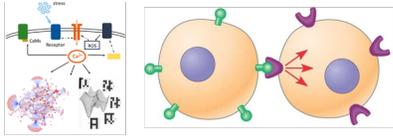

**(A')** 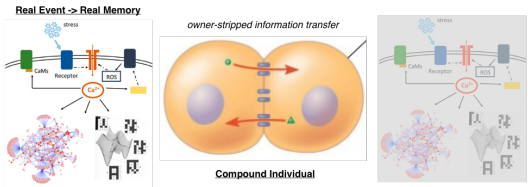

**(B)** 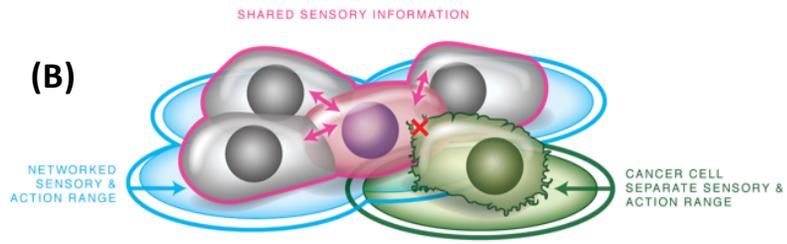

**(C)** 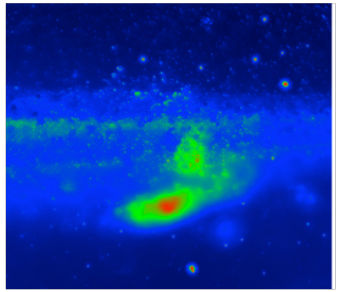

**(D)** 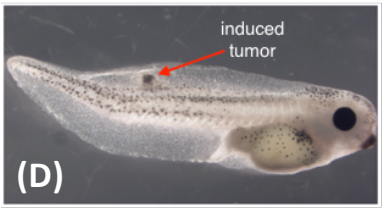

**(E)** 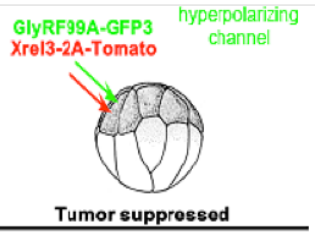

**(F)** 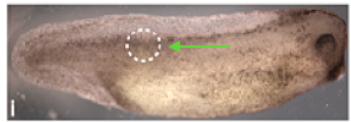

**(G)** 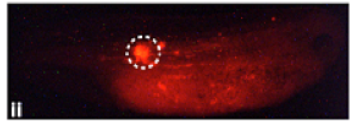

Levin, Figure 9

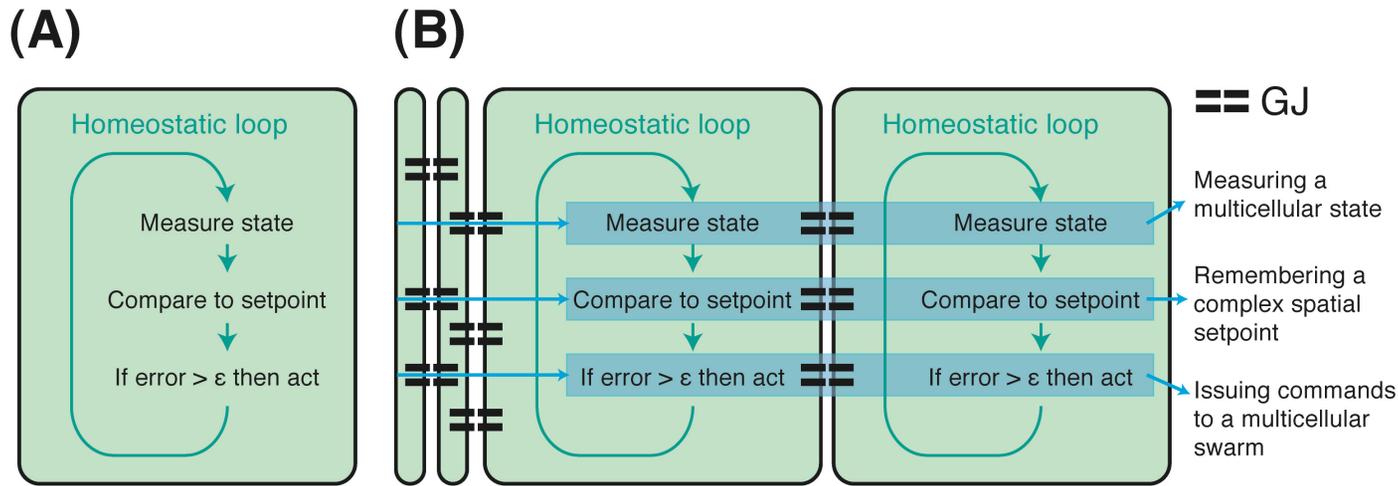
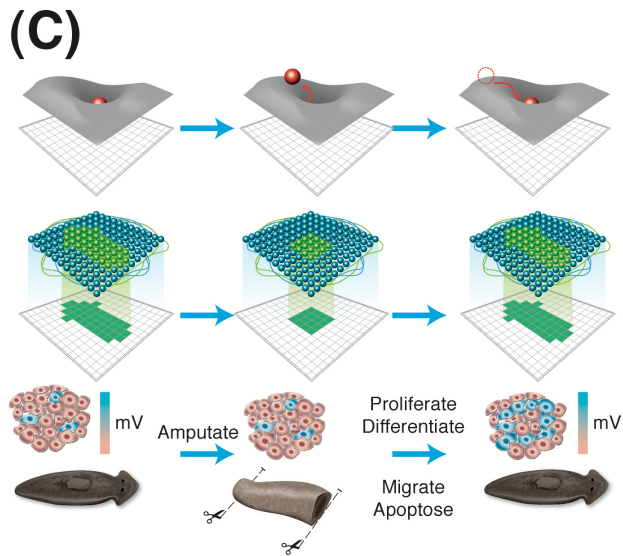
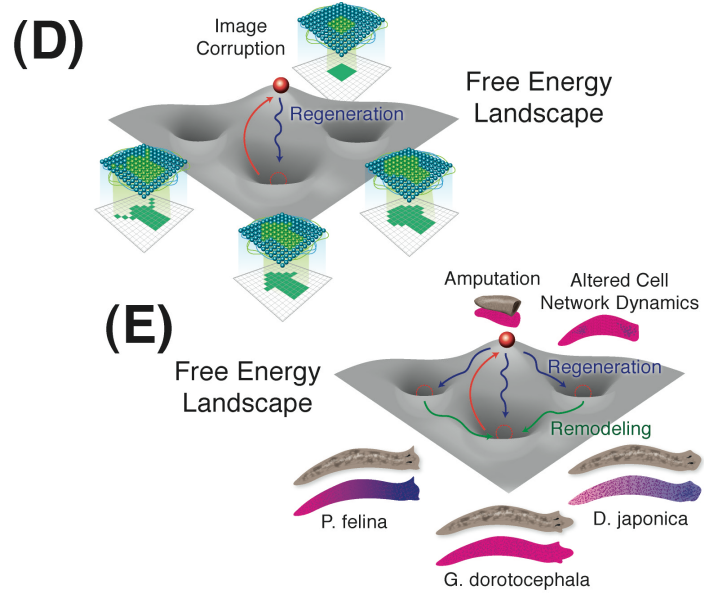

Levin, Figure 10

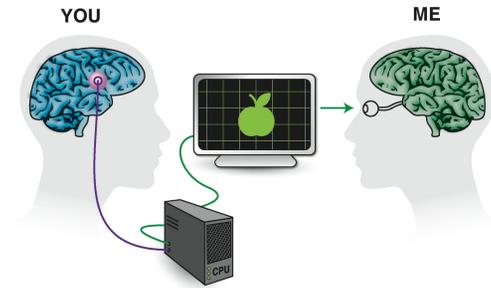
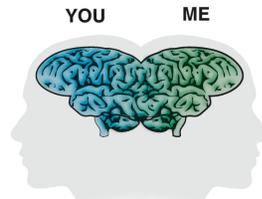
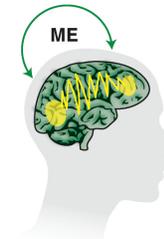
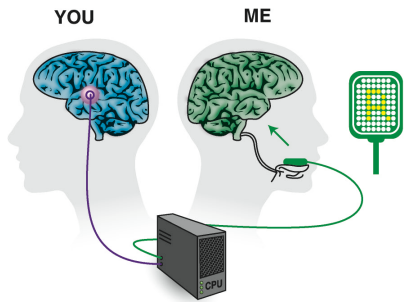
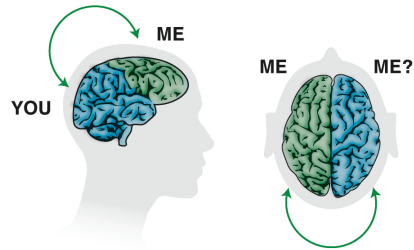

Levin, Figure 11